\documentclass{aa}  
\usepackage{amssymb}

\usepackage{comment}
\usepackage{graphicx}

\usepackage{txfonts}
\usepackage[colorlinks=true,linkcolor=red, citecolor=blue]{hyperref}
\usepackage{fontawesome}

\begin{document} 

\newcommand{\Rb}{{R_{\rm b}}}

\titlerunning{Inference of reionization bubbles around galaxy groups}
\title{Mapping reionization bubbles in JWST era II.\\Inferring the position and characteristic size of individual bubbles}

\author{Ivan Nikolić
        \inst{1}
        \and
        Andrei Mesinger\inst{1}
        \and 
        Charlotte A. Mason \inst{2, 3}
        \and
        Ting-Yi Lu \inst{2, 3}
      \and
    Mengtao Tang\inst{5}
    \and
    David Prelogović \inst{1, 4}
    \and
    Samuel Gagnon-Hartman\inst{1}
    \and
    Daniel P. Stark\inst{5}
}

   \institute{Scuola Normale Superiore, Piazza dei Cavalieri 7, 56125 Pisa, PI, Italy
   \and 
    Cosmic Dawn Center (DAWN)
    \and
    Niels Bohr Institute, University of Copenhagen, Jagtvej 128, 2200 Copenhagen N, Denmark
    \and
    Scuola Internazionale Superiore di Studi Avanzati (SISSA), Via Bonomea 265, 34136 Trieste, Italy
    \and
    Steward Observatory, University of Arizona, 933 N Cherry Ave, Tucson, AZ 85721, USA
   }

   \date{Received; accepted}

  \abstract
   {
   The James Webb Space Telescope ({\it JWST}) is discovering an increasing number of galaxies well into the early stages of the epoch of reionization (EoR).  Many of these galaxies are clustered with strong Lyman-alpha (Ly$\alpha$) emission, which indicates surrounding cosmic HII regions that would facilitate Ly$\alpha$ transmission through the intergalactic medium (IGM).  Detecting these HII bubbles would allow us to connect their growth to the properties of the galaxies inside them.
   We developed a new forward-modeling framework to estimate the size of the local HII region and its location based on Ly$\alpha$ spectra of galaxy groups in the early stages of the EoR.  Our model uses the complementary information provided by neighboring sightlines through the IGM.  Our forward models sample the main sources of uncertainty, including (i) the global neutral fraction, (ii) the EoR morphology, (iii) emergent Ly$\alpha$ emission, and (iv) NIRSpec instrument noise.
   Depending on the availability of complementary nebular lines, $\sim$ 0.006 -- 0.01 galaxies per cMpc$^3$ are required for a confidence of $\gtrsim$95\%  that the location and size of the HII bubble as recovered by our method is accurate to within $\sim$ 1 comoving Mpc. This approximately corresponds to some dozens of galaxies at $z\sim7$--8 in an $\sim$2x2 tiled pointing with {\it JWST} NIRSpec. A sample like this can be achieved with a targeted survey that is complete down to $M_{\rm UV}^{\rm min}\lesssim$ -19 -- -17, depending on the overdensity of the field.
   We tested our method on 3D EoR simulations and on mispecified equivalent-width distributions. We accurately recovered the HII region that surrounded the targeted galaxy groups in both cases.
  }


   \maketitle

\section{Introduction}

The epoch of reionization (EoR) marks an important milestone in the evolution of the Universe. UV radiation from the first clustered galaxies reionized their surrounding intergalactic medium (IGM).  These HII bubbles expanded, percolated, and eventually permeated all of space, completing the final phase change of our Universe.  The timing and morphology of the EoR tell us which galaxies caused this, and it describes the role of the IGM clumps that regulated the end stages \citep[e.g.,][]{McQuinn2007, Sobacchi2014}.

Lyman-$\alpha$ emission from galaxies is an especially useful tool for studying the early stages of the EoR, when HII regions were relatively small such that the neutral IGM left a strong imprint via damping wing absorption \citep[e.g., see review in][]{Dijsktra2014}.
It is a common approach to estimate the mean neutral fraction of the IGM using a statistically large sample of galaxies (e.g., \citealt{Mesinger2008b, Stark2010,Mesinger2015, Mason2018, Jung2020, Bolan2022, Jones2024, Nakane2024}). Galaxies reside in biased regions of the IGM, however, and the connection of the corresponding damping wing signature to the mean neutral fraction depends very strongly on the model \citep[e.g.,][]{Mesinger2008a, Lu2023}.
In contrast, unbiased probes such as the Lyman-$\alpha$ forest are sourced from representatively-large volumes of the IGM, and can already tightly constrain the mean neutral fraction during the latter half of the EoR \citep{Qin2021, Bosman2022, Qin2024b}.

In addition to estimating the global neutral fraction from the IGM Ly$\alpha$ damping wings, the presence (or absence) of an individual HII region surrounding an observed group of galaxies could be inferred \citep{Tilvi2020, Endsley2022, Jung2022, Hayes2023}. This might allow us to connect the growth of the local HII region to the properties of the galaxies inside it.

Several such estimates of HII bubble sizes would allow us to understand which type of galaxies drove reionization \citep[e.g., faint/bright;][]{McQuinn2007, Mesinger2016}, and this would be possible well before the advent of tomographic 21 cm maps with the Square Kilometer Array (SKA). 
Fortunately, the James Webb Space Telescope (JWST) is providing spectra from an ever-increasing number of galaxy groups at high redshift that can be used for this purpose \citep[e.g.,][]{Saxena2023, Witstok2024, Tang2023, Tang2024b, Chen2024, Umeda2024, Napolitano2024}

However, the interpretation of these observations has so far been fairly approximate. An IGM damping wing in each galaxy is estimated independently of its neighbors.  This wastes invaluable information because neighboring galaxies provide complimentary sightlines into the local EoR morphology.  The result is that the studies focusing on individual galaxies generally only predict lower limits for the radii of local HII regions. Furthermore, studies tend to ignore one or more than one important source of stochasticity in the EoR morphology, intrinsic galaxy emission, and/or telescope noise. 

We develop a new framework to  infer the local HII region size and location from Ly$\alpha$ observations of a galaxy group. Our formalism accounts for the relative position of each galaxy with respect to the host and nearby HII bubbles by creating self-consistent forward models of JWST/NIRSpec spectra for each galaxy.  We account for the relevant sources of uncertainty (and/or stochasticity), including (i) the IGM mean neutral fraction, $\bar{x}_{\rm HI}$; (ii) the EoR morphology, given $\bar{x}_{\rm HI}$ ; (iii) the emergent Ly$\alpha$ emission, given the observed UV magnitudes; and (iv) the NIRSpec instrument noise.
Unlike many previous studies, we made no assumptions about the unknown relative contribution of observed versus unobserved galaxies to the growth of the local HII region. We quantify how many galaxies are required to robustly detect individual ionized regions with an accuracy of $\lesssim10\%$ in their inferred location and characteristic size during the early stages of EoR.  This work is a companion to  \citet{Lu2024}, in which we presented a complementary formalism to detect edges of ionized regions using empirically calibrated relations.

This paper is structured as follows. In Section~\ref{sec:models} we present our forward-modeling pipeline for Ly$\alpha$ galaxy spectra during the EoR. We introduce the procedure with which we infer the size and location of the surrounding HII region in Section~\ref{sec:transmissions_more}. We apply our framework to mock data and show our main results in Section~\ref{sec:results}. We build further confidence by performing out-of-distribution tests in Section~\ref{sec:oud_tests}. In Section~\ref{sec:discussion} we quantify observational requirements for detecting individual HII regions, and we conclude in Section~\ref{sec:conclusion}. All quantities are presented in comoving units unless stated otherwise.  We assume a standard $\Lambda$CDM cosmology
($\Omega_\textrm{m}, \Omega_\textrm{b}, \Omega_{\Lambda}, h, \sigma_8, n_\textrm{s} = 0.310, 0.049, 0.689, 0.677, 0.81, 0.963$), with parameters consistent with the latest estimates from \citet{Planck2020}.  All quantities are quoted in comoving units and are evaluated in the rest frame unless stated otherwise.

\begin{figure*}[h]
    \centering
    \includegraphics[width=\linewidth]{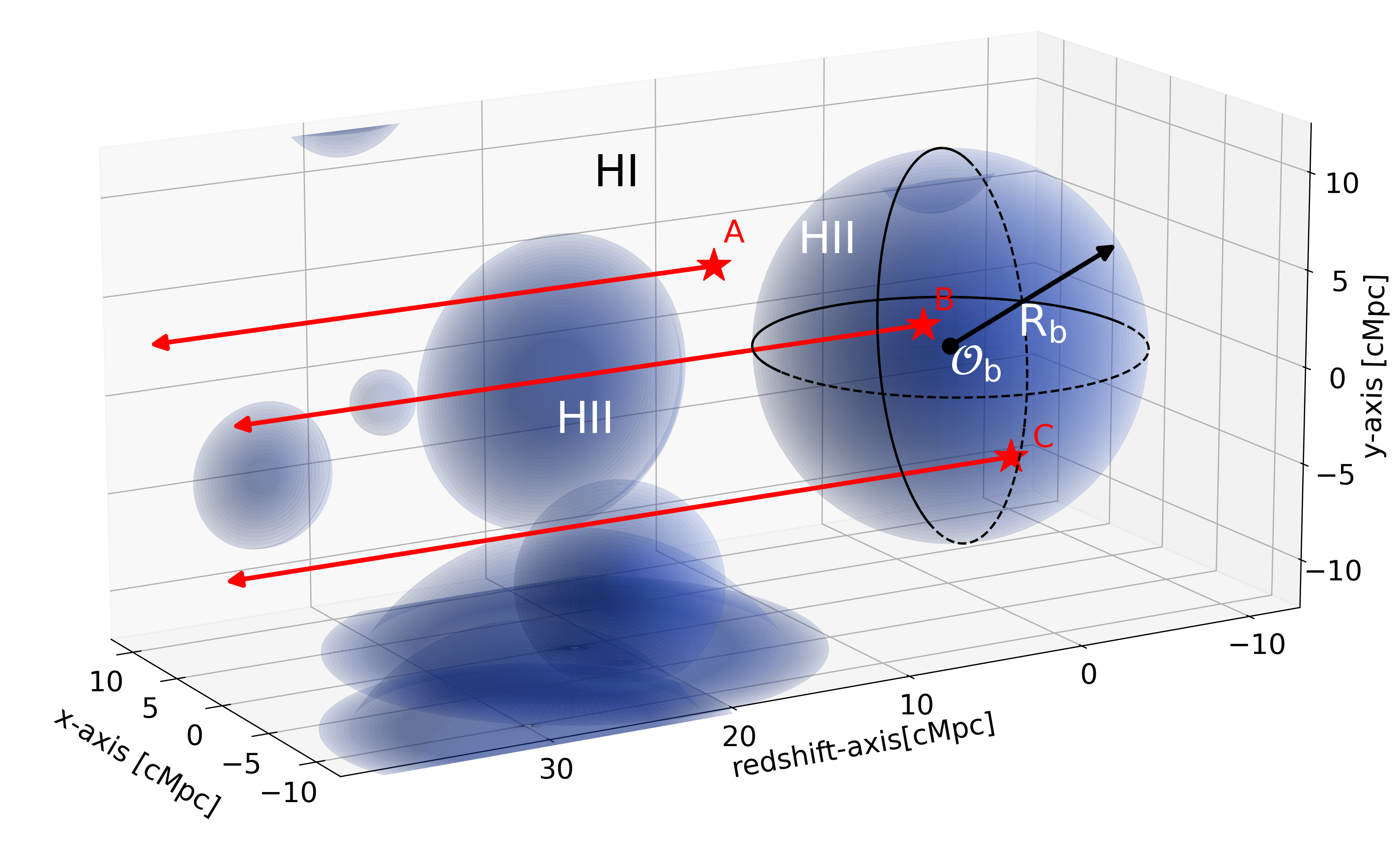}
    \caption{Schematic of our framework. Cosmic HII regions are shown in blue, embedded in an otherwise neutral IGM.    We observe Lyman-$\alpha$ spectra from a group of galaxies (three example galaxies are denoted A, B, and C), and we wish to infer the central HII bubble (characterized by a sphere with radius $R_{\rm b}$ and center $\mathcal{O}_{\rm b}$).  Our framework combines the complimentary information provided by neighboring sightlines toward the galaxies (red arrows) to account for the main sources of stochasticity.}
    \label{fig:setup}
\end{figure*}

\section{Observing Lyman-$\alpha$ spectra from galaxies during the EoR}
\label{sec:models}

Our fiducial setup is shown in Fig.~\ref{fig:setup}.  An HII region in an observed galaxy field is characterized as a sphere, with a center location ($\mathcal{O}_{\rm b}$) and characteristic radius ($\Rb$). This is the local or target HII region whose properties we aim to infer.  Nearby ionized regions are also present.  Observed galaxies can be located both inside and outside HII regions; we denote three such galaxies with A, B, and C and highlight their sightlines toward the observer with red arrows.

Specifically, we wished to determine the  conditional probability of the HII bubble center, $\mathcal{O}_{\rm b}$, and radius, $\Rb$, based on the observed Ly$\alpha$ spectra of $N_{\rm gal}$ galaxies in a field with a central redshift $z$,

\begin{equation}
    \mathcal{P}(\mathcal{O}_{\rm b}, R_{\rm b} | \boldsymbol{x}^i, f_{\alpha}^i(\lambda), M_{\rm UV}^i, z) ~, ~ i\in [1, N_{\rm gal}].
\end{equation}
\noindent Here, $\boldsymbol{x}^i$, $M_{\rm UV}^i$, and $f_{\alpha}^i(\lambda)$ are vectors of the Cartesian coordinates of the galaxies, the UV magnitudes, and the observed Ly$\alpha$ spectra.
For each galaxy, the observed spectrum in the rest frame can be written as
\begin{equation}
    f_{\alpha}(\lambda) = L_{\alpha} J (\lambda)  e^{-\tau_{\rm EoR}(\lambda)} + \mathcal{N}(\lambda),
    \label{eq:flux_eq}
\end{equation}
where $L_{\alpha}$ is the emergent\footnote{Throughout we use the term "emergent" to refer to quantities  describing values escaping from the galaxy into the IGM. Therefore the emergent amplitude, $L_\alpha$, and profile, $J(\lambda)$, are determined by Lyman-$\alpha$ radiative transfer through the interstellar medium and the circumgalactic medium (e.g. \citealt{Neufeld1990, Laursen2011}).  We did not model the details of this radiative transfer, but instead relied on empirical relations based on post-EoR observations to determine the conditional distributions of $L_\alpha$ and $J(\lambda)$.} Lyman-$\alpha$ luminosity of a galaxy, $J(\lambda)$ is the normalized, emergent  Lyman-$\alpha$ profile, $\exp[-\tau_{\rm EoR}(\lambda)]$ accounts for IGM attenuation, and $\mathcal{N}(\lambda)$ is the spectrograph noise. 

In the schematic shown in Fig. \ref{fig:setup}, galaxy A lies outside of an ionized bubble, and its Lyman-$\alpha$ flux will therefore be strongly attenuated by the neutral IGM (i.e., it will have a large $\tau_{\rm EoR}(\lambda)$).  Lyman-$\alpha$ flux from galaxy A is only expected when it has a high emergent luminosity, $L_{\alpha}$, {\it and} when its Ly$\alpha$ profile, $J(\lambda)$, is strongly redshifted from the systemic $z$ (e.g., \citealt{Dijsktra2014}).  Galaxy B is close to the center of the local HII bubble and will have (on average, see the right panel of Fig. \ref{fig:mean-transmission}), which shows the lowest Ly$\alpha$ damping wing attenuation from the patchy EoR).  
The observed flux depends on all of the terms in eq. (\ref{eq:flux_eq}), however, each of which can have a sizable stochasticity.  Below, we detail how we compute each of these terms.

\begin{figure}
    \centering
    \includegraphics[width=\columnwidth]{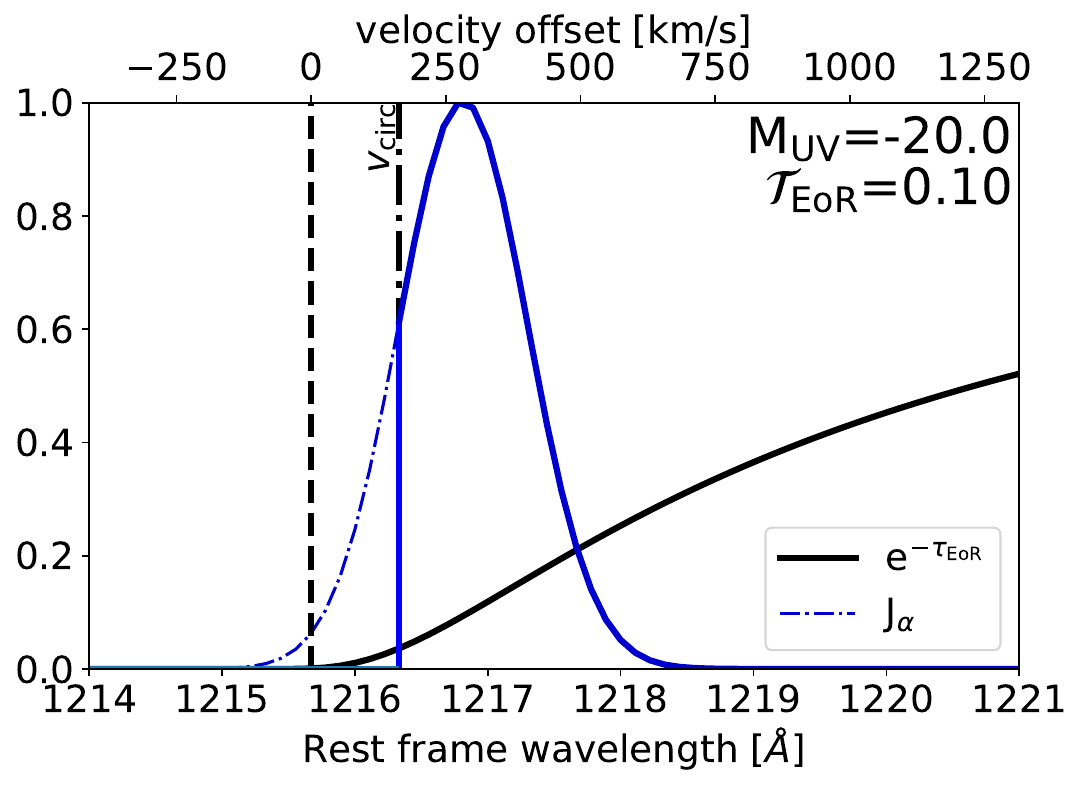}
    \caption{Emergent Lyman-alpha line and IGM opacity as a function of wavelength. The solid blue line represents the normalized Lyman-$\alpha$ emergent profile, and the dot-dashed blue line demarcates the flux absorbed by the circumgalactic medium blueward of the circular velocity of the galaxy. The solid black line illustrates an  IGM damping wing attenuation profile, taken from a random sightline at $\bar{x}_{\rm HI} =0.65$ and $R_b=10$cMpc (see text for details).  The  Lyman-alpha transmission integrated over all wavelengths for this example would be $\mathcal{T}_{\rm EoR}\equiv \int \textrm{d}\lambda J(\lambda)e^{-\tau_{\rm EoR}(\lambda)}=0.10$.}
    \label{fig:example}
\end{figure}

\subsection{Emergent Lyman-$\alpha$ profile}
\label{sec:conditional_J}

\begin{figure}
    \centering
    \includegraphics[width=\columnwidth]{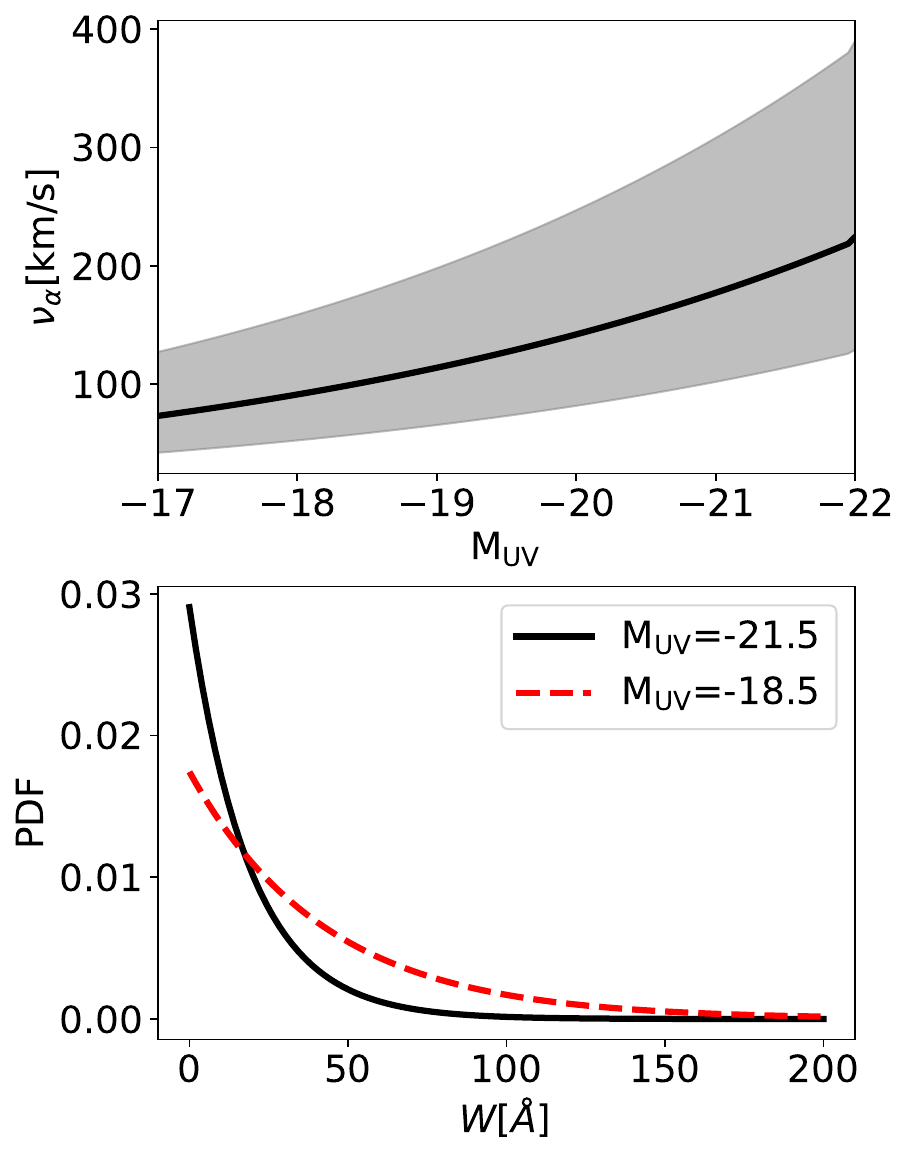}
    \caption{\textit{Upper panel:} Mean (black line) and $1\sigma$ (shaded region) of the velocity offset distribution as a function of UV magnitude (Eq.~\ref{eq:P_delta}). \textit{Lower panel:} PDF of the Lyman-alpha rest-frame equivalent widths. The dashed black and red curves represent the distribution of the equivalent widths ($W$) for Lyman-$\alpha$ emitters (Eq.~\ref{eq:distr_EW}). The equivalent widths of the nonemitters are described by a delta function at $W=0\AA$ (not shown) normalized such that the integral of the PDF is $1$. The distribution is shown for M$_{\rm UV} = -18.5$ and $-21.5$}
    \label{fig:delta_v_EW}
\end{figure}

We started with the  Lyman-alpha profile emerging into the IGM, $J(\lambda)$, normalized to integrate to unity.
In order to escape from the ISM of the galaxy, Lyman-$\alpha$ photons must diffuse spectrally, which leads to a double-peaked Lyman-alpha shape \citep{Neufeld1990, Hu2023, Hutter2023, AlmadaMonter2024}. Because the line is resonant, the blue peak is generally absorbed even by the ionized IGM at $z\gtrsim5$ (but see \citealt{Meyer2021} for some putative rare counterexamples). Following \citet{Mason2018}, we modeled the remaining red peak as a Gaussian,

\begin{equation}
    J(\lambda) = \frac{2}{\nu_{\alpha}}\sqrt{\frac{\ln{2}}{\pi}} \exp{\left(-\frac{(\Delta {\rm v} - v_{\alpha})}{v_{\alpha}^2 / \left(2 \ln{2}\right)}^2\right)},
    \label{eq:J-shape}
\end{equation}
where $\nu_{\alpha}$ represents the velocity offset from the systemic velocity of the center of the line, and $\Delta {\rm v} = ((\lambda - \lambda_{\alpha}) c / \lambda_{\alpha})$ is the velocity difference from the resonant wavelength of the line, $\lambda_{\alpha}=1215.16\AA$. As in \citet{Mason2018}, we assumed for simplicity that the FWHM of the line is equal to the velocity of the offset \citep[e.g.,][]{Verhamme2018}.   Although these profiles are motivated by lower-redshift observations (e.g., \citealt{Yamada2012, Orlitova2018, Hu2023}), we note that our framework will easily accommodate any distribution for $J(\lambda)$ when we have better models for the emergent spectra.  We also assumed that all Lyman-alpha photons with a velocity offset below the circular velocity, $v_{\rm circ}$, of the host halo are absorbed by the CGM \citep{Dijsktra2011,Laursen2011}. In Fig.~\ref{fig:example} we show an example of the emergent profile in blue for a galaxy with an UV magnitude $M_{\rm UV}=-20.0$ and a velocity offset $v_{\alpha}=270$km/s, with a $v_{\rm circ} = 160$km/s.

The PDF of the emergent velocity offset is well described by a log-normal distribution \citep{Steidel2014, DeBarros2017, Stark2017, Mason2018}:
\begin{equation}
    \mathcal{P}(v_{\alpha} | M_{\textrm{UV}}) = \frac{1}{\sqrt{2 \pi} \ln{10} \  v_{\alpha} \ \sigma_{\rm v}} \exp{\left(- \frac{(\log_{10}{v_\alpha} - \overline{v_{\alpha}}(M_{\rm UV}))^2}{2 \sigma_{\rm v}^2}\right)},
    \label{eq:P_delta}
\end{equation}
where the mean velocity offset is correlated with the UV magnitude \citep[][though see \cite{Bolan2024}]{Stark2017},

\begin{equation}
    \log_{10}{\overline{v_{\alpha}}}(M_{\textrm{UV}},z) = 0.32 \gamma (M_{\textrm{UV}} + 20.0 + 0.26 z) + 2.34
\end{equation}

and $\sigma_{v} = 0.24$, $\gamma=-0.3$ for $M_{\rm UV}\geq-20.0 - 0.26 z$, and $\gamma=-0.7$ otherwise. We show this distribution in the upper panel of Fig.~\ref{fig:delta_v_EW}. Although there is some indication of a mild redshift evolution in this distribution (e.g., \citealt{Tang2024b, Witstok2024b}), we show in Section~\ref{sec:oud_tests} that our results are insensitive to such changes.

\subsection{Emergent Lyman-$\alpha$ luminosity}
\label{sec:la_luminosity}

The absolute normalization of the profile discussed above (i.e., the emergent Lyman-$\alpha$ luminosity $L_\alpha$) is generally defined via the so-called rest-frame equivalent width \citep[e.g.,][]{Dijkstra2012},

\begin{equation}
\label{eq:EW}
W = \frac{L_{\alpha}}{L_{1500,\nu}} \frac{\lambda_{\alpha}}{\nu_{\lambda}}\left(\frac{\lambda_{\rm UV}}{\lambda_{\alpha}}\right)^{\beta+2},
\end{equation}
where $L_{1500,\nu}[\rm erg/s/Hz]$ is the specific UV luminosity evaluated at $1500\AA$ obtained from the continuum flux $f_{1500, \nu}$: $L_{1500,\nu} = 4\pi d_{\rm L}^2 f_{1500}$ where $ f_{1500}$ is given in units of $[\rm erg/s/cm^2/Hz]$, where $d_{\rm L}$ is the luminosity distance to the source, $\nu_{\lambda}=2.47 \times 10^{15}$Hz, $\lambda_{\alpha} = 1215.67\AA$, $\lambda_{\rm UV}=1500\AA$ is the rest-frame wavelength at which the UV magnitude is measured and $\beta$ is the UV slope (which we assumed to be $\beta=-2.0$ for simplicity).

For  galaxies at $z<6$, where we expect $\tau_{\rm EoR}$ to be negligible, \citet{Mason2018} found the following fit based on data from \citet{DeBarros2017}:

\begin{equation}
    p_6(W|M_{\rm UV}) = \frac{A(M_{\rm UV})}{W_{\rm c}(M_{\rm UV})} e^{-\frac{W}{W_{\rm c}(M_{\rm UV)}}} H(W) + [1-A(M_{\rm UV})] \delta(W),
    \label{eq:distr_EW}
\end{equation}

\noindent where $A(M_{\rm UV})$ is the fraction of intrinsic emitters for a given $M_{\rm UV}$, and $W_{\rm c}(M_{\rm UV})$ is the characteristic scale of the distribution, which is anticorrelated with $M_{\rm UV}$. $H(W)$ is the Heaviside step function, and $\delta(W)$ is a Dirac delta function. We used the following fit, as in \citet{Mason2018}: $A = 0.65 + 0.1 \tanh{[3(M_{\rm UV} + 20.75)]}$ and $W_{\rm c} = 31 + 12 \tanh{[4(M_{\rm UV} + 20.25)]} \AA$. The distribution of emergent equivalent widths is shown in the lower panel of Figure \ref{fig:delta_v_EW} for two UV magnitudes. 
 We note that our framework can easily accommodate different EW distributions (e.g., \citealt{Treu2012, Lu2024, Tang2024b}). 
We show in Section~\ref{sec:oud_tests} that our results are not sensitive to the choice of distribution shape, however.

\subsection{IGM damping wing absorption}
\label{sec:bubble_distribution}

\begin{figure}
    \centering
    \includegraphics[width=\columnwidth]{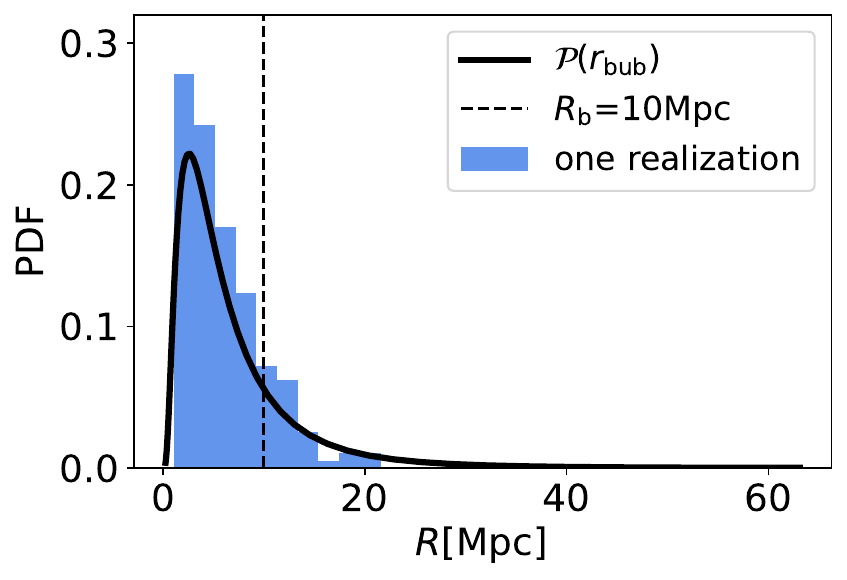}
    \caption{Distribution of HII region radii used in our overlapping spheres algorithm for the large-scale EoR morphology (see text for details). The vertical dashed line marks $R_{\rm b} = 10$cMpc which is our fiducial size for the central ionized region whose properties we aim to infer (see Fig. 1).}
    \label{fig:bubble_distribution}
\end{figure}

\begin{figure*}
    \centering\includegraphics[width=\textwidth]{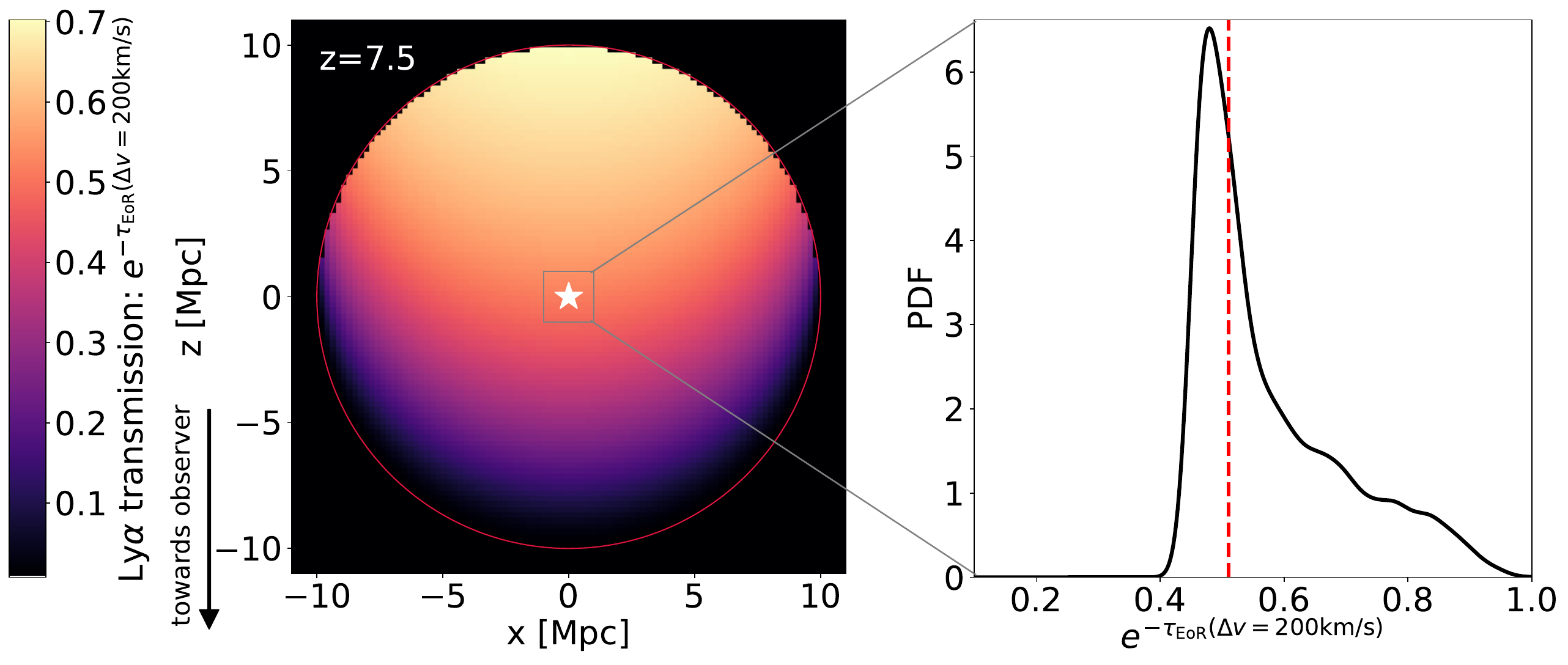}
    \caption{{\it Left panel}: {\it Mean} Ly$\alpha$ transmission evaluated at $\Delta {\rm v} = +200$km/s  ($e^{-\tau_{\rm EoR}}(\Delta{\rm v} = +200)$km/s) as a function of the position inside an ionized bubble.  The observer is located toward the bottom of the figure.  We computed the mean transmission by averaging over $10000$ realizations of EoR morphologies, given an assumed neutral fraction of $\bar{x}_{\rm HI} = 0.65$ at $z=7.5$ (see text for details). {\it Right panel}: {\it Distribution} of Ly$\alpha$ transmission at $\Delta {\rm v} = +200$km/s from these realizations, evaluated at a fixed position inside the bubble denoted by the white star (see text for details).  The mean is marked by the vertical dashed line.}
    \label{fig:mean-transmission}
\end{figure*}

During the EoR, the damping wing absorption from the residual HI patches along the line of sight can strongly attenuate the Ly$\alpha$ line (c.f. the $\exp[-\tau_{\rm IGM}]$ curve in Fig. \ref{fig:example}).  The optical depth of the damping wing is mostly sensitive to the distance to the nearest neutral HI patch (e.g., \citealt{MiraldaEscude1998}). For this reason, we are able to infer the size of the local HII bubble in this work (see also the complementary empirical approach in \citealt{Lu2024} based on empirical $\tau_{\rm IGM} - R_b$ relations from an EoR simulation).

Nevertheless, the surrounding EoR morphology beyond the local HII region does contribute to the total $\tau_{\rm IGM}$ as an additional source of scatter (e.g. \citealt{Mesinger2008a}).

We generated an EoR morphology {\it at a given $\bar{x}_{\rm HI}$} by placing overlapping spherical HII regions with a radius distribution given by (cf. Fig. ~\ref{fig:setup})
\begin{equation}
\label{eq:bubsize}
    \mathcal{P}(r_{\rm bub} | \bar{x}_{\rm HI}) = \frac{1} {\sqrt{2 \pi} \sigma_{\rm bub}} \cdot \exp{\left(-\frac{(\log{r_{\rm bub}} - \mu_{\rm bub})^2}{\sigma_{\rm bub}^2}\right)}.
\end{equation}

We sampled from the above distribution of radii and randomly chose the center location. We stopped when the volume-filling factor of ionized regions reached the desired value, $\bar{x}_{\rm HI}$.  Although this is simplistic, the overlapping ionized spheres resulted in a similar EoR morphology as is seen in cosmological radiative transfer simulations \citep[e.g.][]{Zahn2011, Mesinger2011, Ghara2018,  Doussot2022}. We did not assumed that we know the true value $\bar{x}_{\rm HI}^{\rm true}$ a priori; instead, we sampled a prior distribution of $\bar{x}_{\rm HI}$ from complimentary observations while performing inference (see Section~\ref{sec:transmissions_more} for more details).

For simplicity, we ignored in this proof-of-concept work the $\bar{x}_{\rm HI}$ dependence of the bubble size distribution in eq. (\ref{eq:bubsize}) and took constant values of $\mu_{\rm bub} = \log(5 {\rm cMpc})$ and $\sigma_{\rm bub} = 0.5$.  These choices in our algorithm roughly reproduce the ionized bubble scales seen in simulations during the early stages of the EoR (e.g., \citealt{Mesinger2007, Giri2018, Lu2024, Doussot2022, Neyer2024}). The $r_{\rm bub}$ from Eq.~\ref{eq:bubsize} does not directly translate into any of the metrics that are commonly used to characterize the EoR morphology \citep[e.g.][]{Lin2016, Giri2018}, and comparisons must therefore be made a posteriori.  In future work, we will calibrate Eq. (\ref{eq:bubsize}) to EoR simulations and will also condition it on the matter field to account for the (modest) bias of HII regions at early times \citep[e.g. Fig. 12 in][]{Sobacchi2014}.  We show our assumed bubble size distribution  in Fig.~\ref{fig:bubble_distribution} together with one realization in a $200\times80\times80$ cMpc$^3$ volume.

For a given realization of the EoR morphology and galaxy field (c.f. Fig.~\ref{fig:setup}), we computed the optical depth of the damping wing by casting rays from the galaxy locations and summing the contributions from the optical depth from all HI patches along the LOS (e.g. \citealt{MiraldaEscude1998}),

\begin{equation}
    \tau_{\rm EoR} (\lambda_{\rm em}) = \frac{\tau_{\rm GP} R_{\alpha}}{\pi} \sum_i x_{\rm H,i} \left(\frac{1+z_{b,i}}{1+ z_{\lambda}}\right)^{3/2} \left[ I \left(\frac{1+z_{b,i}}{1+ z_{\lambda}}\right) - I\left(\frac{1+z_{e,i}}{1+ z_{\lambda}}\right)\right],
    \label{eq:tau_ME98}
\end{equation}

where $z_{\lambda} = \frac{\lambda_{\rm em}}{\lambda_{\alpha}} (1+z_{\rm obs}) - 1$, $\lambda_{\rm em}$ is the wavelength at which we evaluated the optical depth, $\lambda_{\alpha} = 1215.67\AA$ is the Lyman-$\alpha$ resonant wavelength, and $z_{\rm obs}$ is the redshift of the emitting galaxy. $\tau_{\rm GP}\sim 7.16\cdot 10^5 \left((1+z_{\rm obs})/10\right)^{3/2}$ is the Gunn-Peterson optical depth, $R_{\alpha} = \frac{\Lambda}{4\pi\nu_{\alpha}}$, $\Lambda = 6.25 \cdot 10^8 \textrm{s}^{-1}$ is the decay constant, and $\nu_{\lambda} = 2.47 \cdot 10^{15}$Hz is the Lyman-$\alpha$ resonant frequency. In the above equation, $I(x)$ is given by

\begin{equation}
    I(x) \equiv \frac{x^{9/2}}{1+x} + \frac{9}{7}x^{7/2} + \frac{9}{5}x^{5/2} + 3x^{3/2} + 9x^{1/2} - \ln\left|\frac{1+x^{1/2}}{1-x^{1/2}}\right|.
\end{equation}

The summation accounts for every neutral patch encountered along the LOS, with a given patch, $i$, extending from $z_{b,i}$ to $z_{e,i}$. We assumed that ionized patches have no neutral hydrogen atoms and therefore do not contribute to the attenuation.

In the left panel of Fig~\ref{fig:mean-transmission}, we show the {\it mean} IGM transmission, evaluated at $\Delta {\rm v}=+200$km/s redward of the systemic redshift, as a function of the position inside the central HII bubble of $R_{\rm b} = 10$ cMpc.  The observer is located toward the bottom of the figure.  This mean transmission was computed by averaging over 10000 realizations of EoR morphologies outside the central bubble, assuming $\bar{x}_{\rm HI}=0.65$ (e.g., one such realization is shown in Fig. \ref{fig:setup}).  As expected, there is a clear trend of increased transmission for galaxies located at the far end of the central HII bubble.  The mean transmission is a function of the distance of the galaxy to the bubble edge in the direction toward the observer.  In \citealt{Lu2024} we used these trends to define empirical edge-detection algorithms.

At every location in the bubble, however, there is sizable sightline-to-sightline scatter in the IGM transmission.  We quantify this in the right panel of Fig. \ref{fig:mean-transmission}, where we  show the transmission PDF constructed from the 10000 realizations of the EoR morphology external to the central bubble.  The sightlines we used to compute this PDF originated in the location marked by the white star in the left panel.  The PDF is quite broad and asymmetric (see also Fig. 2 in \citealt{Mesinger2008b} and \citealt{Lu2023}). While it is difficult for the IGM to completely attenuate Ly$\alpha$ for a galaxy that is located at the center of this bubble, some morphologies can result in large stretches of ionized IGM in the direction of the observer, driving a high-transmision tail in the PDF.   The width of this PDF highlights that it is import to account for stochasticity in the EoR morphology when galaxy Lyman-$\alpha$ observations are interpreted (e.g., \citealt{Mesinger2008a, Mesinger2015, Mason2018, Bruton2023, Keating2024, Lu2024}).

\subsection{Including NIRSpec noise}
\label{sec:noise}

We took NIRSpec on the JWST for our fiducial spectrograph. It already measures Ly$\alpha$ spectra from galaxy groups during the EoR (e.g., \citealt{Tang2024, Witstok2024}).

To forward-model NIRSpec observations, we binned the observed flux $
    f_{\rm \alpha} (\lambda)= L_{\alpha} J(\lambda) e^{-\tau_{\rm EoR}(\lambda)}
$ to $R\sim2700$ (the high resolution of NIRSpec) and then added Gaussian noise $\mathcal{N}$ to each spectral bin with a standard deviation of $\sigma(\mathcal{N})=2\times10^{-20}$ergs$^{-1}$cm$^{-2}\AA^{-1}$.  This level of noise can be obtained with a few hours of integration on NIRSpec \citep{Bunker2023, Saxena2023, Tang2024}, and it corresponds to an uncertainty on the integrated flux $\left(\mathcal{F}_{\rm int} \equiv \int f_{\rm \alpha}(\lambda) \textrm{d}\lambda\right)$ of $\sigma(\mathcal{N_{\rm int}}) = 1 \times 10^{-19}$ergs$^{-1}$cm$^{-2}$ (estimated assuming that the emission line is spectrally unresolved). This can be further translated into $5\sigma$ limiting equivalent widths of $W=25\AA$ ($W=60\AA$) for $M_{\rm UV}=-18$ ($M_{\rm UV}=-17$). The recovery is worse for shallower observations, but our results did not improve significantly for integrations deeper than this fiducial value.

We rebinned the spectra to lower values of $R$ and tested the inference with a coarser resolution.  
By performing additional binning, we lose some information on the observed line profile \citep[e.g.,][]{Byrohl2020} but lower the dimensionality of our likelihood (see Section~\ref{sec:transmissions_more}).  In future work, we will explore more sophisticated inference approaches that can scale to high-dimensional likelihoods \citep[][]{Cranmer2020, Montel2024}.  We empirically settled on rest frame $\Delta\lambda \sim 1\AA$ as our bin width (i.e., $R\sim 1000$), which corresponds to the medium-resolution NIRSpec grating \citep[][see Section~\ref{sec:transmissions_more}]{Jakobsen2022}.

\section{Inferring the local HII bubble}
\label{sec:transmissions_more}

\begin{figure*}
    \centering
    \includegraphics[width=\linewidth]{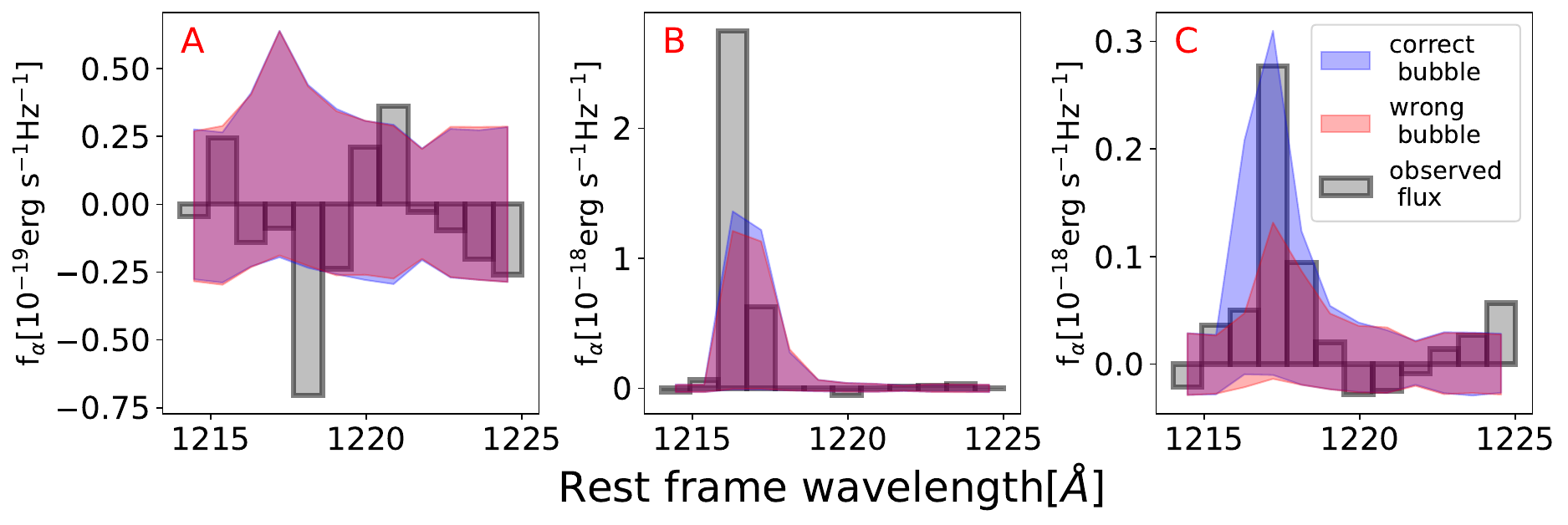}
    \caption{Observed flux ({\it gray}) corresponding to the galaxies A, B, and C from the mock observation shown in Fig. \ref{fig:setup}. We show the 68\% C.L. of the likelihood assuming the correct HII bubble location and radius in blue, $(R_{\rm b}, \mathcal{O}_{\rm b})  = (R_{\rm b}^{\rm true}, \mathcal{O}_{\rm b}^{\rm true})$.  We show the 68\% C.L. of the flux likelihood assuming the correct HII bubble location, but a slightly smaller radius in red, $R_{\rm b} = 0.8 R_{\rm b}^{\rm true}$.}
    \label{fig:flux_fws}
\end{figure*}

With the above framework, we created mock observations and the corresponding forward models by sampling each of the terms in Eq. \ref{eq:flux_eq}.  We detail this procedure below.

\subsection{Mock observations}

We first constructed a mock observation of $N_{\rm gal}$ galaxies in a survey volume of $V_{\rm survey}=$ 20x20x20 cMpc$^3$ ($\sim \ 7$'$\ \times 7$'$\ \times \ \Delta z=0.07$) at $z=7.5$.  This FoV is roughly motivated by {\it JWST} (corresponding to roughly four NIRSpec pointings), although in practice, the forward-modeled volume should be tailored to the specific observation that is interpreted.  We placed a bubble with radius $R_b=$ 10 cMpc at the (arbitrarily chosen) center of the volume and constructed the surrounding EoR morphology out to distances of 200 cMpc\footnote{Neutral IGM at larger distances contributes a negligible amount to the total attenuation because the damping wing profile is steep \citep{Mesinger2008a}.} using the prescription from Sect.~\ref{sec:bubble_distribution} and assuming $\bar{x}_{\rm HI}=0.65$.  We demonstrate below that our results are insensitive to these fiducial choices.

We assigned random locations to the galaxies, $\boldsymbol{x}^i$ with $i\in[1, N_{\rm gal}]$, using rejection sampling to ensure that 75\% of the galaxies are inside HII regions on average. This number was motivated by the analysis reported by \citep{Lu2023}, who showed that $\sim25\%$ galaxies are found outside of ionized bubbles for $M_{\rm UV}>-18$ at $\overline{x}_{\rm H}=0.7$ (their Fig.3).  This is a very approximate way of accounting for galaxy bias because galaxies and HII regions are both correlated to the large-scale matter field.\footnote{This is a reasonable approximation, as evidenced by our results in Section~\ref{sec:21cmfast_check}, where we apply our framework to simulations that account for galaxy and HII region bias self-consistently.  Observations of galaxies have demonstrated that the bias dominates clustering at larger scales (some dozen cMpc), while the smaller scales that are relevant for this work are dominated by Poisson noise \citep[][Davies et al. in prep]{Bhowmick2018, Jespersen2024}.}

We generated UV magnitudes for each galaxy by sampling the UV luminosity function (LF) from \citet{Park2019} down to a magnitude limit of $M_{\rm UV}=-18.0$.  Each galaxy was then assigned an emergent emission profile according to the procedure in the previous section, which was attenuated by its sightline through the realization of the EoR morphology.  Finally, a noise realization was added to the binned flux to create a mock spectrum for each galaxy (c.f. Eq. \ref{eq:flux_eq}).

\subsection{Maximum likelihood estimate of the bubble size and location}

\begin{figure*}
    \centering
    \includegraphics[width=1.0\linewidth]{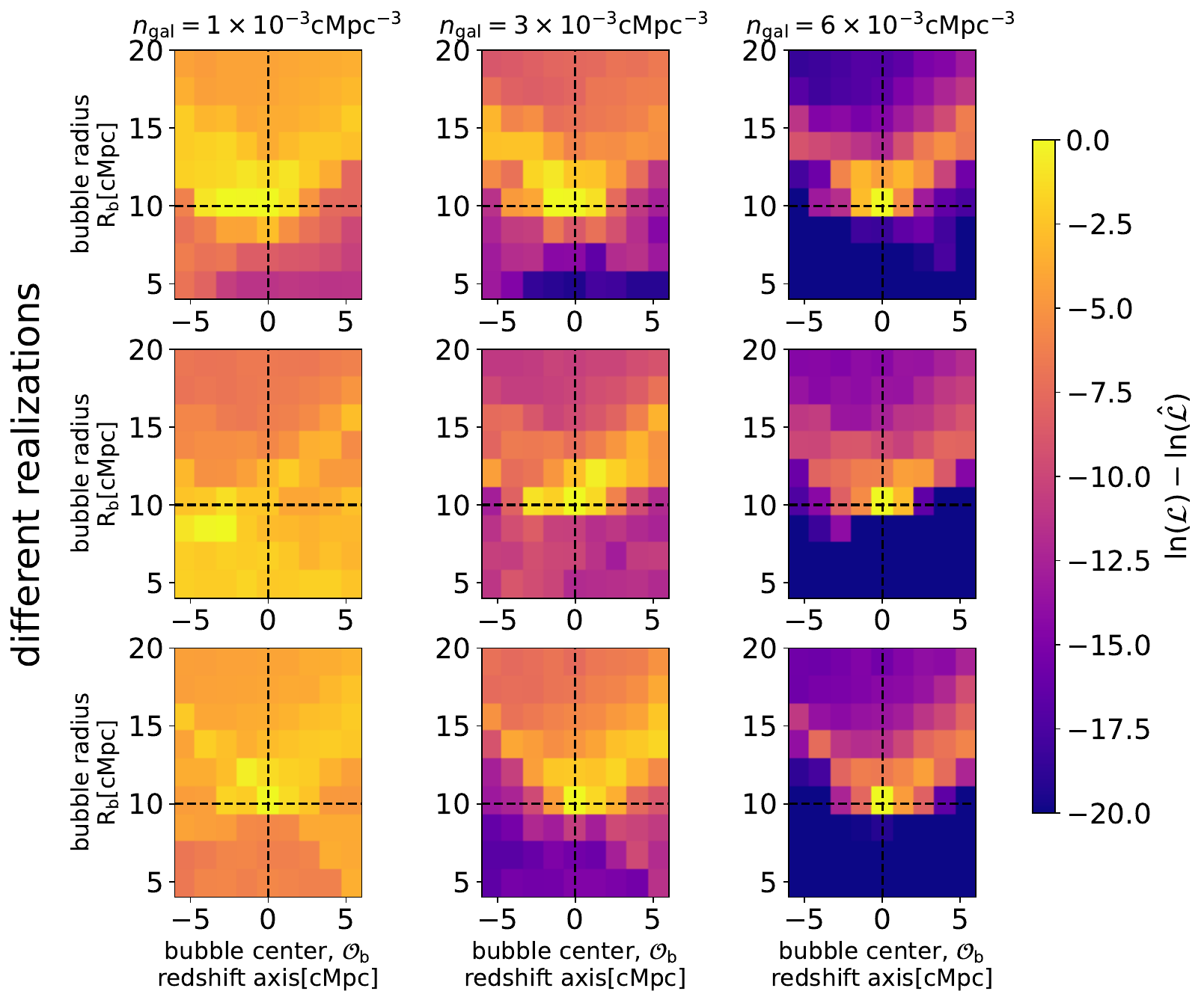}
    \caption{Two-dimensional slices through the log likelihood, normalized to the maximum value in each panel. The vertical and horizontal axis in each panel correspond to the bubble radius, $R_{\rm b}$, and redshift axis of the center, $\mathcal{O}_{\rm b} = (x_b = x_b^{\rm true}, y_b = y_b^{\rm true}, z_b)$.  The true values, $(R^{\rm true}_{\rm b}, z^{\rm true}_b)$ = (10 cMpc, 0 cMpc),  are demarcated with the dashed lines. The mock observation in this example was made at $z=7.5$, assuming $\overline{x}_{\rm H}=0.65$. The different columns represent different numbers of galaxies that were used for inference (8 galaxies on the left, 24 galaxies in the middle, and 48 galaxies on the right). The different rows correspond to different realizations of the EoR morphology and galaxy samples.}
    \label{fig:likelihoods}
\end{figure*}

We then interpreted this mock observation by forward-modeling the observed flux for each galaxy and by varying (i) the position and radius of the central HII bubble; (ii) the surrounding EoR morphology; (iii) the neutral fraction of the Universe (within $\pm 0.1$ of the truth, which we fixed conservatively to wider than current limits \citealt{Qin2024b}); (iv) the emergent Lyman-alpha flux given the observed $M_{\rm UV}$ of the galaxy (i.e., $W$ and $v_\alpha$); and (v) NIRSpec noise realizations.

We then interpreted this mock observation by forward-modeling the observed flux for each galaxy and by varying (i) the position and radius of the central HII bubble; (ii) the surrounding EoR morphology; (iii) the neutral fraction of the Universe (within $\pm 0.1$ of the truth, which we fixed conservatively to wider than current limits \citealt{Qin2024b}); (iv) the emergent Lyman-alpha flux given the observed $M_{\rm UV}$ of the galaxy (i.e., $W$ and $v_\alpha$); and (v) NIRSpec noise realizations.

For each forward model, we computed the likelihood of the mock observation based on the location and radius of the central HII region.  Our sampling procedure effectively marginalized over the unknowns (ii) -- (v) from above.  Because it is numerically challenging to map out the joint likelihood over all of the observed galaxies, we made the simplifying assumption that the likelihood of the observed flux from each galaxy, $f_{\alpha}^i(\lambda)$, is independent from the other galaxies.  This allowed us to write the total likelihood of the observation as a product of the likelihoods of the individual galaxies,
\begin{equation}
\label{eq:likelihood}
 \mathcal{L} = \prod_i^{N_{\rm gal}} \mathcal{L}^i(f_{\alpha}^i(\lambda) ~|~ \mathcal{O}_{\rm b}, R_{\rm b},  \boldsymbol{x}^i, M^i_{\rm UV}, z).
\end{equation}
\noindent  While this assumption is clearly incorrect, we only present results here in terms of the maximum likelihood, $\hat{\mathcal{L}}$.  We demonstrate below that Eq. (\ref{eq:likelihood}) provides an unbiased estimate of  $\hat{\mathcal{L}}$.\footnote{Ideally, we would want to map out the full posterior:
\begin{equation}
\label{eq:posterior}
  \nonumber  \mathcal{P}(\mathcal{O}_{\rm b}, R_{\rm b} | \boldsymbol{x}, f_{\alpha}(\lambda), M_{\rm UV}, z)  \propto \mathcal{L}(f_{\alpha}(\lambda) | \mathcal{O}_{\rm b}, R_{\rm b},  \boldsymbol{x}, M_{\rm UV}, z) ~ \pi(\mathcal{O}_{\rm b}, R_{\rm b}) ,
\end{equation}
\noindent where $\pi(\mathcal{O}_{\rm b}, R_{\rm b})$ is a prior for the center location and radius of the HII bubble.  However, including the correlations of the (non-Gaussian) likelihoods at the location of every galaxy is analytically not tractable, and would require high dimensional simulation based inference \citep[e.g.,][]{deSanti2023, Lemos2023}.  We save this for future work.
}

It is important to note that {\it we did not assume a Gaussian likelihood for the observed flux at each wavelength}, as is commonly done.  With high S/N spectra, the correlations between flux bins can be significant. We instead directly mapped out the {\it joint PDF of flux over all wavelength bins}, $\mathcal{L}^i(f_{\alpha}^i(\lambda_1), f_{\alpha}^i(\lambda_2), f_{\alpha}^i(\lambda_3), ... ~|~ \mathcal{O}_{\rm b}, R_{\rm b},  \boldsymbol{x}^i, M_{\rm UV}^i, z)$ using a kernel density estimation over the forward-modeled spectra (see Appendix~\ref{sec:appendixA} for details).  This preserved the covariances between the wavelength bins and is commonly known as an implicit likelihood or a simulation-based inference. 

We demonstrate this procedure for the three galaxies shown in Fig. \ref{fig:setup}. For the observed flues from galaxies A, B, and C we show the 68\% C.L. of the likelihood assuming the correct HII bubble location and radius, $(R_{\rm b}, \mathcal{O}_{\rm b})  = (R_{\rm b}^{\rm true}, \mathcal{O}_{\rm b}^{\rm true})$. We also show  the 68\% C.L. of the flux likelihood assuming the correct HII bubble location but a slightly smaller radius, $R_{\rm b} = 0.8 R_{\rm b}^{\rm true}$.  Galaxy A in this mock observation is outside of the central HII bubble, and therefore the true and slightly smaller values of $R_{\rm b}$ result in the same likelihood. Galaxy B is located close to the center of the bubble, so both $R_{\rm b}=R_{\rm b}^{\rm true}$ and $R_{\rm b}=0.8 R_{\rm b}^{\rm true}$ give similar values of  transmission $\exp[-\tau_{\rm EoR}]$ (see Fig.~\ref{fig:mean-transmission}). This results in only a slight preference for $R_{\rm b}=R_{\rm b}^{\rm true}$. On the other hand, galaxy C is located close to the edge of the bubble in Fig.~\ref{fig:setup}. For this galaxy, the observed spectrum in gray is more consistent with the correct likelihood in blue than with the incorrect likelihodd in red.  A smaller bubble would imply a stronger IGM attenuation on average at the location of this galaxy, which makes the observed strong Ly$\alpha$ emission less likely.  In this specific realization, the joint likelihood (Eq. \ref{eq:likelihood}) of the observed fluxes of A, B, and C is twice higher for the correct value of the bubble radius than for the incorrect value. When progressively more galaxies are included, the maximum likelihood increasingly peaks around the true values for the bubble size and location.

We illustrate this explicitly in Figure \ref{fig:likelihoods}, in which we plot 2D slices through the log likelihood.  The log likelihood normalized to the maximum value is indicated with the color bar.  The vertical and horizontal axis in each panel correspond to the sampled bubble radius, $R_{\rm b}$, and to the redshift axis of the center, $\mathcal{O}_{\rm b} = (x_b = x_b^{\rm true}, y_b = y_b^{\rm true}, z_b)$.  The true values, $(R^{\rm true}_{\rm b}, z^{\rm true}_b)$ = (10 cMpc, 0 cMpc),  are demarcated with the dashed lines. The columns correspond to increasing number density of observed galaxies ({\it left to right}), and the rows correspond to different realizations of forward models.

\begin{figure*}
    \centering
    \includegraphics[width=\textwidth]{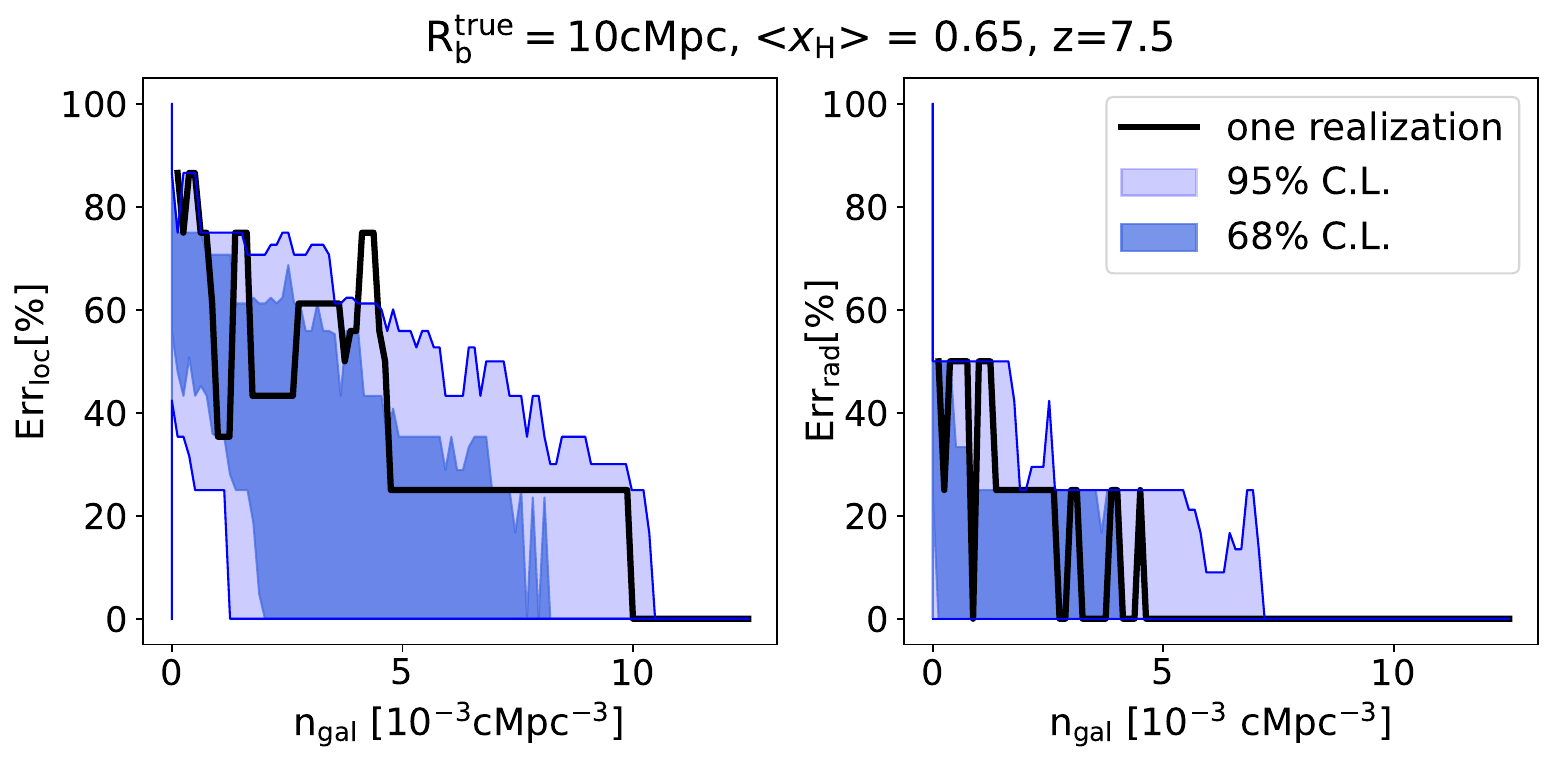}
    \caption{Fractional errors of the maximum likelihood HII bubble center ({\it left}) and radius ({\it right}) as a function of the number density of the observed galaxies.
The black curve corresponds to a single realization of the EoR morphology and galaxy samples.  The light (dark) blue region corresponds to the 95th (68th) percentile of fractional errors obtained from 100 realizations. 
With $n_{\rm gal}\gtrsim 0.01$ galaxies cMpc$^{-3}$, the fractional errors drop below the grid resolution we used to calculate the likelihoods ($1.5$cMpc).}
    \label{fig:summary}
\end{figure*}

Several interesting trends are evident in Figure \ref{fig:likelihoods}.  First, an increasing number of observed galaxies (from left to right in each row) results in an increasingly peaked likelihood, and the maximum settles on the true values for the bubble radius and location.  Different realizations of the EoR morphology give different likelihood distributions when a small number of galaxies is observed. As the number density of the observed galaxies is increased, however, the likelihood distributions between different realizations converge (i.e., all sources of stochasticity are averaged out).

The redshift of the bubble center and its radius are also degenerate. Because the transmission is mostly determined by the distance to the bubble edge along the line of sight, a movement of the bubble center farther away from the observer at a fixed radius is roughly degenerate with a decrease in the radius at a fixed center location.  This degeneracy is mitigated by a larger number of sightlines to observed galaxies, which allows us to constrain the curvature radius of the bubble.

\section{How many galaxies do we need to confidently infer the local HII bubble?}
\label{sec:results}

In the previous section, we demonstrated that our framework gives an unbiased estimate of the HII region size and location when it is applied to a large galaxy sample. We quantify below just how large this galaxy sample needs to be in order for us to be confident in our results.
 
For this purpose, we defined two figures of merit,
\begin{align}
	\textrm{Err}_{\rm loc} &= \left| \vec{\hat{\mathcal{O}}_{\rm b}}(n_{\rm gal}) - \vec{\mathcal{O}_{\rm b}^{\rm true}} \right|  / R_{\rm b}^{\rm true}\\
   \nonumber &= \sqrt{(\hat{x}_b - x_b^{\rm true})^2 + (\hat{y}_b - y_b^{\rm true})^2 + (\hat{z}_b - z_b^{\rm true})^2} / R_{\rm b}^{\rm true} \cr\\
	\textrm{Err}_{\rm rad } &= \left| \hat{R}_{\rm b}(n_{\rm gal}) - R_{\rm b}^{\rm true}\right| / R_{\rm b}^{\rm true}. 
\end{align}
\noindent   Here, $\vec{\hat{\mathcal{O}}_{\rm b}}(n_{\rm gal})$ and $\hat{R}_{\rm b}(n_{\rm gal})$ are the maximum likelihood estimates of the HII bubble center and radius computed from a galaxy field with a number density $n_{\rm gal} = N_{\rm gal}/V_{\rm survey}$, and 
 Err$_{\rm loc}$ and Err$_{\rm rad}$ are the corresponding fractional errors (normalized to the true bubble radius).

We calculated the likelihood on the grid, and our framework has therefore found the optimal bubble when the fiducial and inferred locations and radii coincide, that is, Err$_{\rm loc} = $ Err$_{\rm rad}=0$. In this case, the error of the location and radius is below the grid on which we calculated the likelihood ($1.5$cMpc in our fiducial case, corresponding to a $\lesssim15$\% fractional error).

The solid black curve in Fig.~\ref{fig:summary} shows the change in these fractional errors with galaxy number density for a single realization. The realizations of the EoR morphology and observed galaxies are held constant here, with the maximum likelihood computed each time a new observed galaxy is added.   The more galaxies we observe, the smaller the error on our inferred HII bubble location and radius.  The sizable stochasticity in galaxy properties and sightline-to-sightline scatter in the IGM opacity manifest as noise in this evolution, making it nonmonotonic.  Nevertheless, even a single galaxy is able to shrink the fractional error by a factor of two from our prior range, ruling out extreme values.

We repeated this calculation with 100 different realizations of the mock observation (EoR morphology and observed galaxy samples).  In Fig.~\ref{fig:summary} we show the resulting 68\% and 95\% C.L. on the fractional errors as increasingly more galaxies are added to the field. In 68\% (95\%) of the cases, number densities of $7.7\times$ ($10.5\times$) $10^{-3}$cMpc$^{-3}$ are sufficient to obtain an error of $\lesssim15\%$  on the center position.  The corresponding requirements are $4.2\times$ ($7.2\times$) $10^{-3}$cMpc$^{-3}$ for an error of $\lesssim15\%$ on the bubble radius (comparable to \citealt{Lu2024}).  In other words, Ly$\alpha$ spectra from $\sim0.01$ galaxies per cMpc$^3$ should be observed to be $\gtrsim$95\% confident that the HII bubble location and size recovered by our method is accurate at $\lesssim$ 1 cMpc. This approximately corresponds to 80 galaxies in 2x2 tiled pointings with {\it JWST}/NIRSpec.

\begin{figure*}
    \centering
    \includegraphics[width=\textwidth]{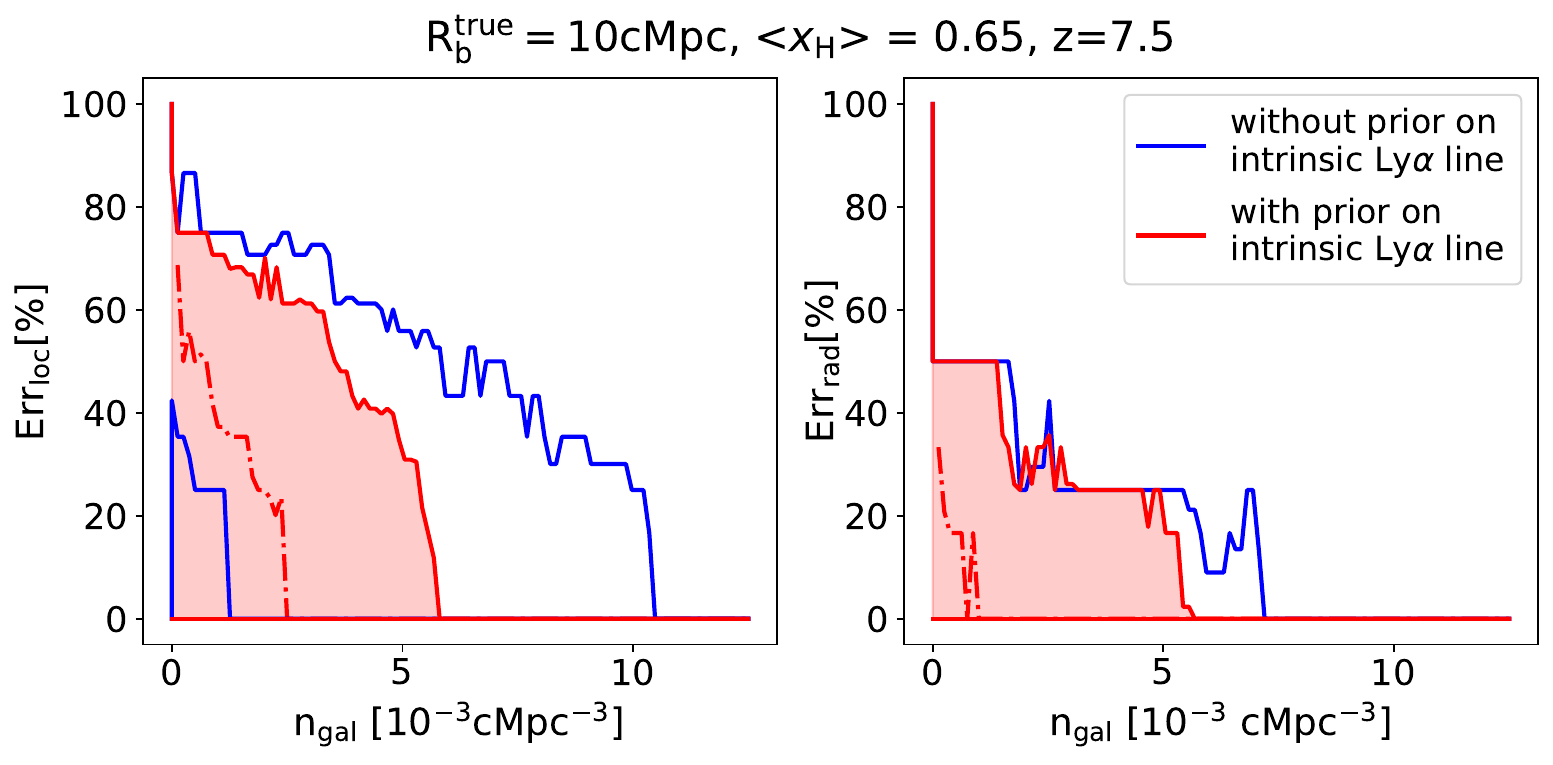}
    \caption{Same as Fig.~\ref{fig:summary}, but with the 95\% C.L. on the fractional errors (red) assuming a prior on the emergent Lyman-alpha emission, motivated by observed Balmer lines (see text for details).
    }
    \label{fig:num_gal_cp}
\end{figure*}

\subsection{Including a prior on the emergent Lyman-$\alpha$}
\label{sec:priors}

We assumed that the emergent Lyman-$\alpha$ emission follows post-EoR empirical relations, as described in Section \ref{sec:conditional_J}. Beyond this, we assumed no prior knowledge on the intrinsic emission of each galaxy. Nebular lines with a lower opacity such as the Balmer lines (H$\alpha$, H$\beta$), however, can provide complimentary estimates of the intrinsic production of Lyman-$\alpha$ photons (e.g., \citealt{Hayes2023b, Saxena2023, Chen2024}).  In this section, we repeat the analysis from above, but with a simple prior on the emergent Ly$\alpha$ emission.

Specifically, we applied a rejection sampling to keep only those forward models for which the transmission was within $20\%$ of the true transmission (or placing an upper bound if $\mathcal{T}_{\rm EoR}\leq 0.2$)\footnote{A detection of Balmer lines places constraints on the production rate of Lyman-alpha photons, which then have to pass through the ISM, CGM, and IGM.  Our distribution of Lyman-$\alpha$ equivalent widths and profile shapes described in Section.~\ref{sec:la_luminosity} effectively already includes radiative transfer and escape through the ISM + CGM, which we assumed does not evolve significantly with redshift at $z\gtrsim6$. Thus, our prior is effectively a prior on the IGM transmission. We leave a self-consistent treatment of the redshift evolution of the ISM+CGM escape fraction and IGM transmission for future work.} Then, we calculated the likelihood in the same way as we did in Section~\ref{sec:transmissions_more}. The width of the prior was motivated by current observations using JWST  (c.f. \citealt{Saxena2023} for H$\beta$ and \citealt{Lin2024} for H$\alpha$ from ground-based instruments). H$\alpha$ observations above $z>7$ are not possible with NIRSpec. For  higher redshifts, we therefore relied on a fainter H$\beta$. Despite this, H$\beta$ is regularly observed in high$-z$ galaxies, and it is therefore reasonable to use the prior for all galaxies \citep[e.g.,][]{Meyer2024}.

In Fig.~\ref{fig:num_gal_cp} we show the 95\% C.L. on the fractional errors with ({\it red}) and without ({\it blue}; same as Fig.~\ref{fig:summary}) the prior information on the emergent Lyman-$\alpha$ emission.
Additional nebular lines to constrain the emergent Lyman-$\alpha$ emission can reduce the number of galaxies that are required to obtain the same constraints by a factor of about two .

\subsection{Results for different bubble sizes}
\label{sec:varying_radii}
\begin{figure}
    \centering
    \includegraphics[width=\linewidth]{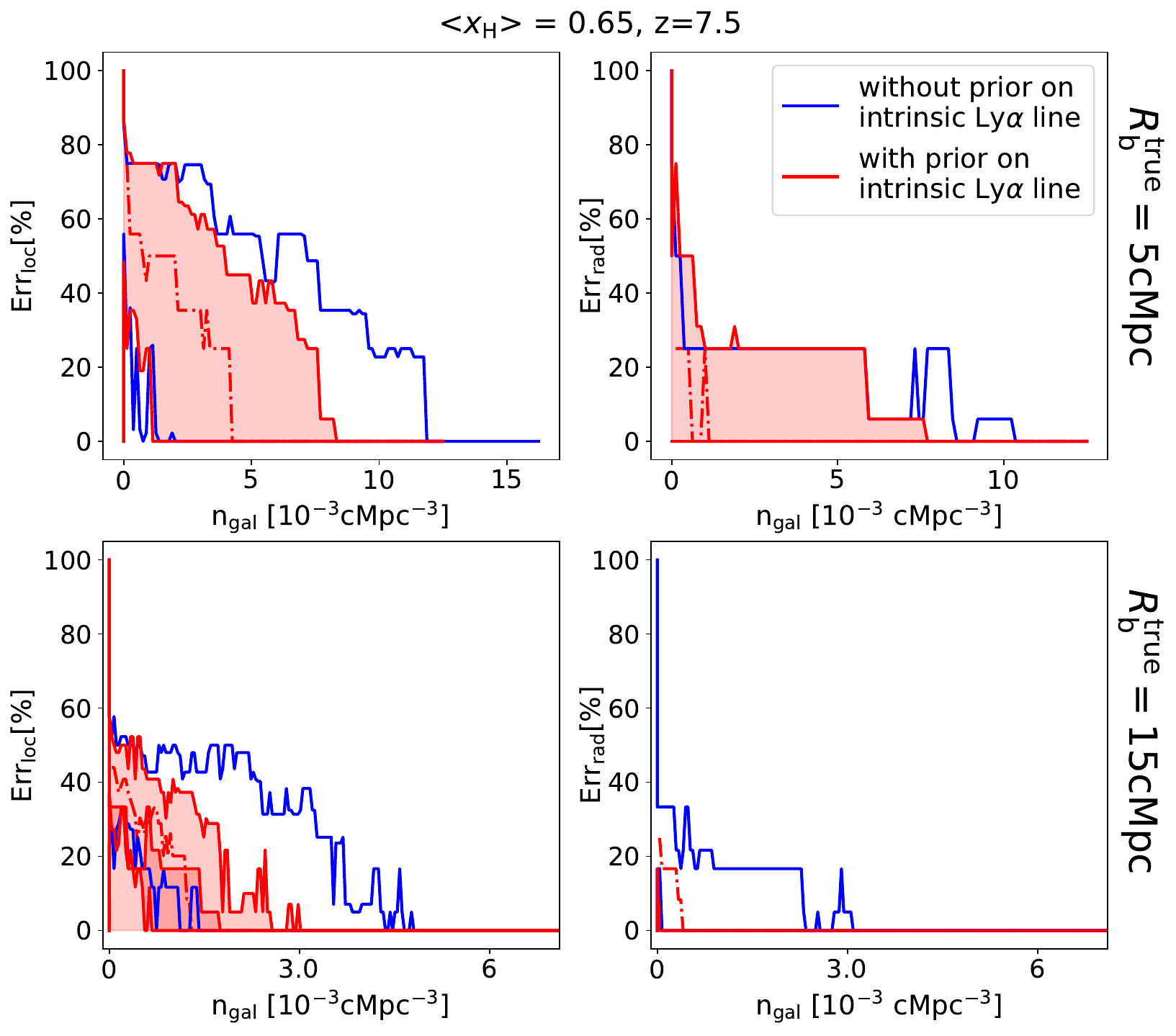}
    \caption{Same as  Fig.~\ref{fig:num_gal_cp}, but assuming $R_{\rm b}^{\rm true} = $ 5 (15) cMpc in the top (bottom) row.  Larger bubbles can be inferred with the same accuracy using a smaller number density of galaxies.}
    \label{fig:R5Mpc}
\end{figure}

Our fiducial choice for the radius of the central HII bubble was $\Rb=10$Mpc, which was motivated by HII bubble sizes in the early stages of the EoR.  
We test the performance of our pipeline for other radii of
$\Rb=5$ and $15$Mpc below.

In the top (bottom) panels of Fig.~\ref{fig:R5Mpc}, we show the analysis with and without the 20\% prior on the  emergent emission, assuming $R_{\rm b}^{\rm true}=5$Mpc ($R_{\rm b}^{\rm true}=15$Mpc).  By comparing to the fiducial results in Fig. \ref{fig:num_gal_cp}, we see that fewer galaxies are required to infer the size and location for larger HII bubbles.  For $R_{\rm b}^{\rm true}=15$ cMpc, about two to three times fewer galaxies are required to constrain the HII bubble with the same accuracy compared to $R_{\rm b}^{\rm true}=5$ cMpc.  The reason most likely is that the larger bubbles allow for a broader range of IGM opacities.  Galaxies inside small bubbles have roughly the same attenuation regardless of their relative location inside the bubble.  In contrast, there is a noticeable difference between the (average) attenuation on the near and far side of the bubble for larger bubbles, which allows us to constrain their geometry more easily (c.f. the left panel of Fig. \ref{fig:mean-transmission}). 

\section{Building confidence in our framework}
\label{sec:oud_tests}
In this section, we explore the dependence of our results on our model assumptions.  To do this, we applied our framework on a different reionization morphology and on different emergent EW distributions.  Even though our framework is a proof of concept, it would help us to build confidence that it is not sensitive to uncertainties in the details of our model if it performed well in these out-of-distribution tests.

\subsection{Different EW distribution}

\begin{figure}
    \centering
    \includegraphics[width=\linewidth]{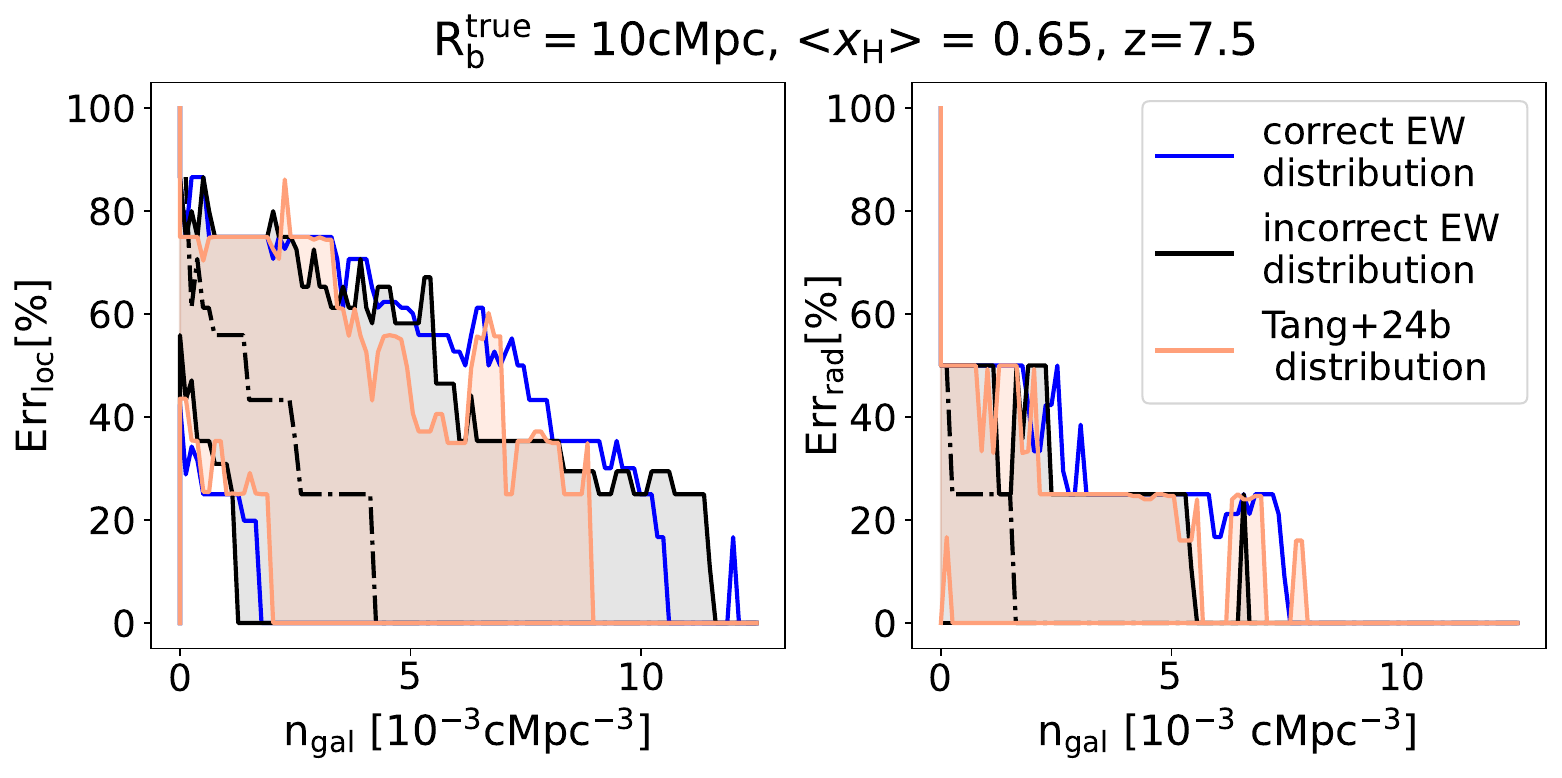}
    \caption{ Same as Fig.~\ref{fig:summary}, but including in gray (orange) the 95\% C.L. on the fractional errors assuming that the EW distribution we used to interpret the mock observation is a Gaussian constructed from the same data (log-normal distribution from \cite{Tang2024b}). In other words, the mock observations and forward models are generated using different EW distributions (see text for details).  The fact that the C.L. demarcated in gray, orange and blue roughly overlap indicates that our results are not sensitive to our choice of the emergent EW distribution.}
    \label{fig:gauss}
\end{figure}

First, we tested the performance of our pipeline assuming the emergent Lyman-$\alpha$ luminosities followed a different EW distribution than the we used in our forward models  (Sec.~\ref{sec:la_luminosity}).
Specifically, we generated mock observations assuming the Gaussian distribution from \citet{Treu2012} (based on \citet{Stark2011}, but with the same parameters as in Section~\ref{sec:la_luminosity}),
\begin{equation}
\begin{aligned}
    p_{6;\rm G}(W|M_{\rm UV}) = \frac{A(M_{\rm UV})}{W_{\rm c}(M_{\rm UV})} \sqrt{\frac{2}{\pi}} & e^{ -\frac{1}{2} \left(\frac{W}{W_{\rm c}\left(M_{\rm UV}\right)}\right)^2} H(W) \\ &+ [1-A(M_{\rm UV})] \delta(W)
    \label{eq:EW_gauss}.
\end{aligned}
\end{equation}
\noindent We then interpreted these mock observations with our fiducial pipeline, which uses the exponential EW distribution from Eq.~\ref{eq:distr_EW}.

\begin{figure*}
    \centering
    \includegraphics[width=\textwidth]{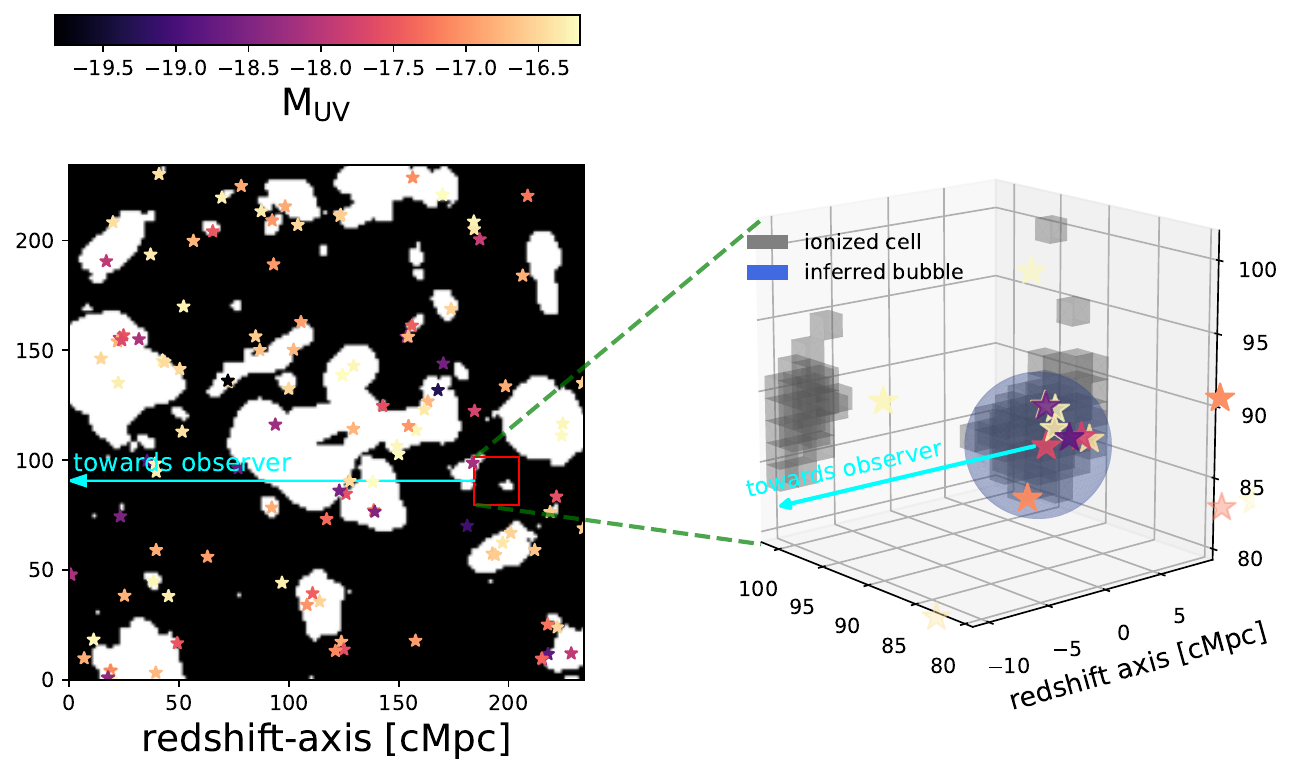}
    \caption{Application of our bubble-finding procedure to a 3D simulation.{\it Left:} 2D slices through the ionization field. Ionized/neutral regions are shown in white/black.  The brightest galaxies in this 1.5 cMpc slice are denoted with stars, colored according to their UV magnitudes.  The line-of-sight direction used to compute the IGM opacity for each observed galaxy is indicated with the arrow. The 20x20x20 cMpc subvolume used for the mock survey is denoted with the red square.{\it Right:} Zoom-in of the mock survey.  A random subsample (to avoid overcrowding) of the observed galaxies is shown with stars.  Ionized voxels are shown in gray.  Our maximum likelihood solution for the central HII bubble is overlaid with the blue sphere.  The reasonable overlap of the central gray blob with the blue sphere shows that our framework is reliable.
    \label{fig:21cmfast_run_069}}
\end{figure*}

The resulting 95\% C.L. on the fractional errors are highlighted in gray in Fig.~\ref{fig:gauss}.  In blue, we show the same 95\% C.L. from Fig.~\ref{fig:summary}, which use the same EW distribution for the forward models and the mock observation.  The fact that the gray and blue regions demarcate roughly the same fractional errors suggests that our analysis is not sensitive to the choice of EW distribution.

As a further test, we repeated the analysis, but with the log-normal distribution from \cite{Tang2024b} that was constructed using completely separate data. The distribution is given with
\begin{equation}
    p(W|M_{\rm UV}) = \frac{1}{\sqrt{2 \pi}\cdot \sigma_{W} \cdot W} \exp\left( -\frac{ \log(W) - \mu_W)^ 2 }{2\cdot \sigma_{W}^2}\right),
\end{equation}

with $\mu_W =\log{(5.0)}$ and $\sigma_W = 1.74$. The results are shown in orange in Fig.~\ref{fig:gauss}. The blue and orange contours again follow each other closely, indicating that our pipeline is robust to the changes in the EW distribution.

\subsection{Demonstration on a 3D reionization simulation}
\label{sec:21cmfast_check}

\begin{figure*}
    \centering
    \includegraphics[width=\textwidth]{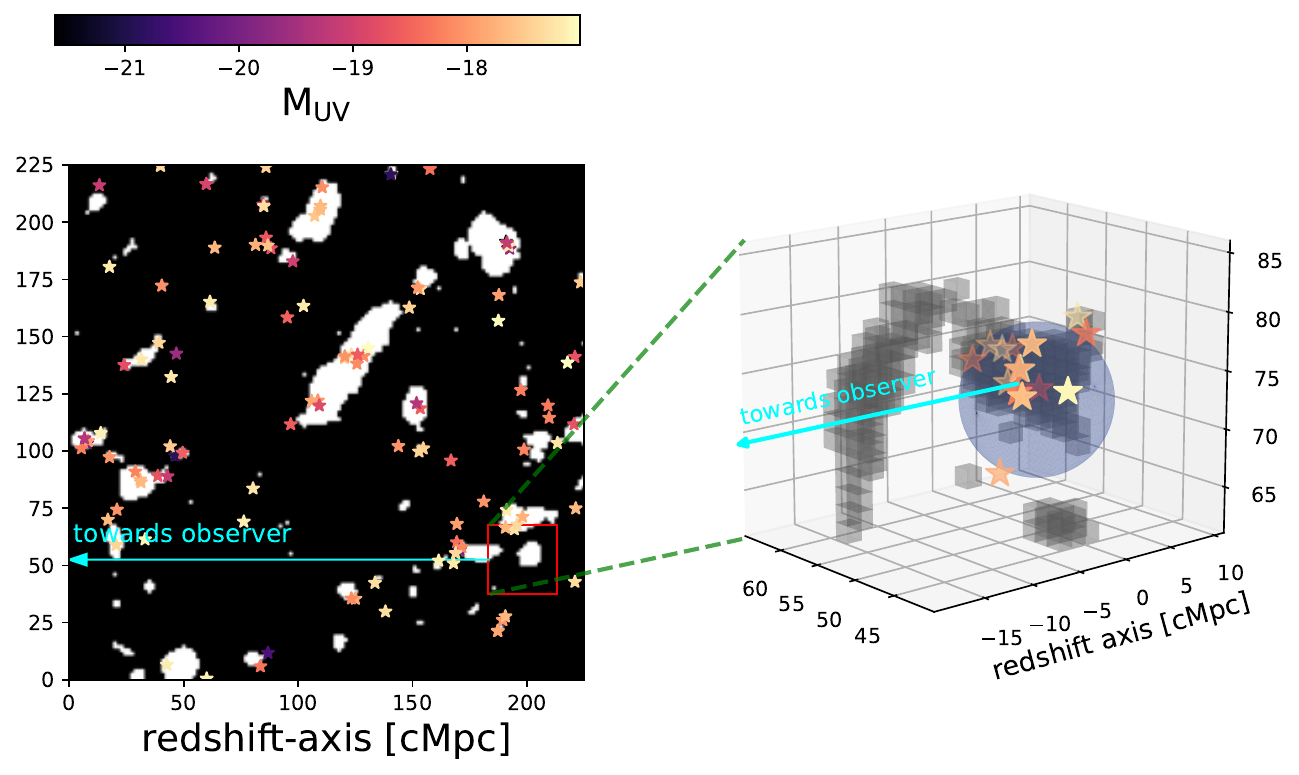}
    \caption{Same as Fig.~\ref{fig:21cmfast_run_069}, but for $\overline{x}_{\rm HI}=0.9$. }
    \label{fig:21cmfast_run_09}
\end{figure*}

Because this work is intended as a proof of concept, we made several simplifying assumptions throughout.  For example, our reionization morphology was generated by overlapping ionized spheres, and we treated the galaxy -- HII bubble bias in a simplified way. A more realistic bias for the galaxies and HII regions could be included straightforward analytically, but the question is whether this is necessary. We tested the performance of our simple model on self-consistent 3D simulations of the reionization.

We applied our framework on the galaxy catalogs and ionization maps from the simulations of \citet{Lu2024}, which were generated with the public {\tt 21cmFAST} code \citep{Mesinger2007, Mesinger2011, Murray2020}.   These simulations capture the complex morphology of the reionization, which is self-consistently generated from the underlying galaxy fields.  We processed the galaxies with the procedure outlined in Section~\ref{sec:models} to create mocks on which we performed the inference.

We used two ionization boxes, one at $\overline{x}_{\rm HI} = 0.69$ and $\overline{x}_{\rm HI} = 0.90$. We selected ionized bubbles and associated galaxies that matched the volumes we used in our fiducial setup, specifically, 20x20x20 cMpc.  When the inference framework is applied to real data, the forward models must be fine-tuned to match the specific details of the survey.  

We illustrate the results in Figures \ref{fig:21cmfast_run_069} and \ref{fig:21cmfast_run_09} for some example surveys at $\overline{x}_{\rm HI}=0.69$ and $\overline{x}_{\rm HI}=0.9$, respectively.
In the left panels, we show 1.5 cMpc thick slices through the simulation boxes. Ionized/neutral regions are shown in white/black.  The brightest galaxies are shown with stars, whose colors correspond to their UV magnitudes.  The line-of-sight direction used to compute the IGM opacity for each observed galaxy is indicated with the arrow.

The volumes of the mock surveys are illustrated with the zoom-ins on the right.  Here, gray cells show the ionized voxels of the simulation, and the blue sphere is the maximum likelihood solution for the central HII region.  We assumed all galaxies inside this 20x20x20  cMpc volume brighter than $M_{\rm UV} < -17.0$ are observed with NIRSpec.  This corresponds to $\sim50 (30)$ galaxies at $\overline{x}_{\rm HI}=0.69$ ($\overline{x}_{\rm HI}=0.9$).

The zoom-ins in these figures show that the inferred HII regions in blue provide a reasonable characterization of the true HII region in gray. This gives us confidence in our simplified treatment of the EoR morphology and the associated spatial correlation with galaxies.  As mentioned above, we will tailor future applications to actual NIRSpec surveys of galaxy groups.

\section{JWST observational requirements}
\label{sec:discussion}

\begin{figure}
    \centering
    \includegraphics[width=\columnwidth]{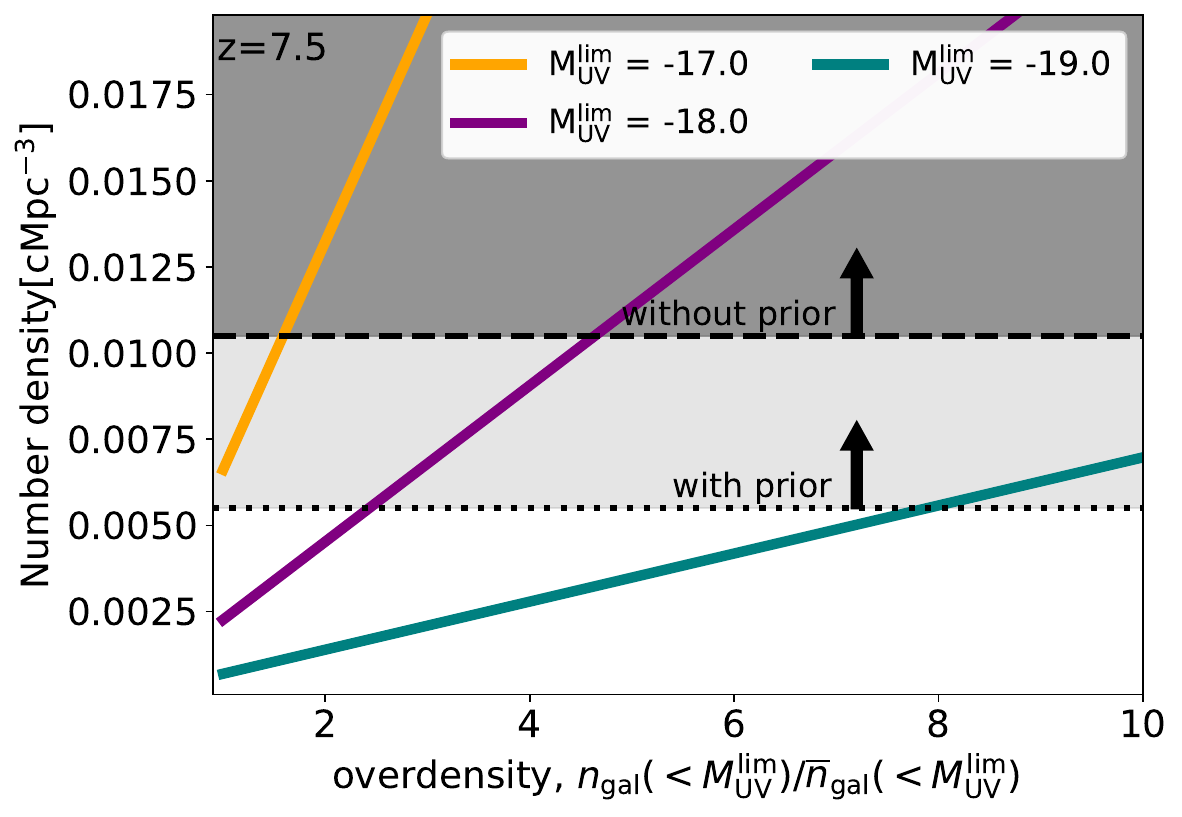}
    \caption{Cumulative number density of galaxies down to various magnitude limits as a function of galaxy overdensity. The dotted (dashed) lines demarcate the minimum number densities we found for an accurate HII bubble recovery, with (without) a prior on the emergent Lyman-$\alpha$ emission.}
    \label{fig:num_gal}
\end{figure}

We currently have several spectroscopically-targeted fields containing groups of galaxies with detection in Lyman-$\alpha$, such as COSMOS \citep{Endsley2022, Witten2024}, EGS \citep{Oesch2015, Zitrin2015, RobertsBorsani2016, Tilvi2020, Leonova2022,Larson2022, Jung2022, Tang2023, Tang2024b, Chen2024, Napolitano2024}, BDF \citep{Castellano2016, Castellano2018}, GOODS-N \citep{Oesch2016, Eisenstein2023, Bunker2023, Tacchella2023}, and GOODS-S \citep{Witstok2024b, Tang2024b}. 
Although the number densities are lower by a factor of a few ($\sim1.0 \times 10^{-3}$cMpc$^{-3}$) than the required values in Sect.~\ref{sec:results}, the data sets are expanding rapidly. Ongoing and proposed programs are extending fields and reach greater depths, which results in larger areas and higher number densities. We quantify the {\it JWST} survey requirements in more detail in order to be able to robustly constrain HII bubbles with our procedure.

In Fig.~\ref{fig:num_gal} we show the number density of galaxies as a function of limiting UV magnitude, $M_{\rm UV}^{\rm lim}$, and galaxy overdensity, $n(< M_{\rm UV}^{\rm lim}/\bar{n}(<M_{\rm UV}^{\rm lim})$.
We plot three curves corresponding to $M_{\rm UV}^{\rm lim}$= -17, -18, and -19. We demarcate in gray the required number density for accurate HII bubble recovery we found in the previous section; the lower/upper range correspond to including/not including a prior on the emergent Ly$\alpha$ emission from Balmer lines (c.f. Section~\ref{sec:results}). 

At the mean number density, we would require the photometric survey to detect galaxies down to $M_{\rm UV}^{\rm lim} \lesssim$ -17.0 -- -17.5. The quoted range spans what can be achieved with or without a prior on the Lyman-$\alpha$ production rate from Balmer lines. In an unlensed field, approximately $600$ hours of integration with NIRCam on the JWST \citep[estimated based on exposure times in][ for 4 pointings]{Morishita2024} would be required to obtain these number densitites, which is achievable, but ambitions.

Several current fields are known to contain overdensities, however. For example, COSMOS  contains a 140 pMpc$^3$ volume that is estimated to be three times overdense at $z\sim 6.8$ \citep{Endsley2022}, while EGS is estimated to contain a $12$pMpc$^3$ volume that is also three times overdense at $z\sim 8.7$ \citep{Zitrin2015, Leonova2022, Larson2022, Tang2023, Chen2024, Lu2023}. Moreover, GOODS-S and GOODS-N contain four and eight times overdensities over volumes that exceed the fiducial volume we used her \citep[62 arcmin$^2$x$\Delta z\sim.2$][]{Tang2024b}. Furthermore, one of the most distant observed LAE is also thought to be located in an overdensity \citep[up to $24\times$ overdense at $z=10.6$ in $2.6$pMpc$^3$ volume;][see also \citealt{Lu2023} for other examples]{Oesch2016, Bunker2023, Tacchella2023}. Observing a field with an overdensity of three (eight) would require photometric completeness down to $M_{\rm UV}^{\rm lim} \lesssim$ -18.5 -- -18 (-19 -- -18.5).  This would require only 50 (4.2) hours of integration with NIRCam. We assumed that the emergent Ly-$\alpha$ properties remain the same depending on the overdensity (i.e., there are no environment dependences in Sections~\ref{sec:conditional_J} and \ref{sec:la_luminosity}).

These candidates then require spectroscopic follow-up. The best candidate for a spectroscopic confirmation is the [OIII]5007$\AA$ emission line. The [OIII] emission line is routinely observed with the JWST \citep[e.g.,][]{Endsley2023, Endsley2024, Meyer2024}. We estimated the required exposure time for detecting [OIII] line by assuming the $M_{\rm UV}$ dependent distribution of the equivalent widths from \citet{Endsley2024}. For this distribution, the equivalent-width limit for detecting $90\%$ of the galaxies at $M_{\rm UV}=-18.0$ is $120\AA$. Eighteen hours per pointing are required to obtain this equivalent width (at $5\sigma$) with the G395M filter. The number density requirement (and thus exposure time) would increase if we were not to detect H$\beta$, as required to place a prior on the intrinsic Lyman-$\alpha$ emission.  It would also increase if the [OIII] distribution were lower at lower magnitudes. 

Following this,  $23$ hours with G140M per pointing are required to detect Lyman $\alpha$ at a noise level of $5\times10^{-19}$erg s$^{-1}$cm$^{-2}$ (corresponding to a detection of a line of W$\approx25\AA$). These numbers are comparable to existing large JWST surveys \citep{Eisenstein2023}. 

We also note that the volume of the potential {\it JWST} surveys used for this analysis has a direct impact on the sizes of the HII regions that can be inferred (see also \citealt{Lu2024}). Tiled surveys of the same depth but with larger volumes could hope to detect correspondingly larger HII regions.  The fact that larger HII regions occur later in the EoR, with correspondingly higher galaxy number densities, might mitigate the required integration time.  We postpone a more systematic survey design to subsequent work in which we apply our framework to simulated 3D EoR light cones.

Another interesting approach might be to couple the framework from \citet{Lu2024} with the framework outlined here. The large-scale empirical  method from \citet{Lu2024} could isolate interesting subvolumes that could then be analyzed with the quantitative inference approach presented here. Finally, the pipeline we developed here could be used to target multiple fields and can be used to infer the sizes of multiple ionized bubbles, which would help us to determine the bubble size distribution. Constraining it would in turn allow us determine the sources that ionized the Universe.

\section{Conclusions}
\label{sec:conclusion}
The morphology of the reionization tells us which galaxies dominated the epoch of cosmic reionization.  Individual bubbles surrounding groups of galaxies encode information on the contribution of unseen, faint sources and allow us to correlate the properties of bubbles and the galaxies they host.

We built a framework to use Lyman-$\alpha$ observations from {\it JWST} NIRspec to constrain the ionization morphology around a group of galaxies. Our framework used for the first time complementary information from sightlines to neighboring galaxies, and it sampled all important sources of stochasticity to robustly place constraints on the size and position of local HII bubbles.

We found that Ly$\alpha$ spectra from $\sim0.01$ galaxies per cMpc$^3$ are required for a confidence of $\gtrsim$95\% that the HII bubble location and size recovered by our method is accurate to better than $\lesssim$ 1.5 cMpc. This approximately corresponds to 80 galaxies in 2x2 tiled pointings with {\it JWST}/NIRSpec.
These requirements can be reduced by using additional nebular lines (e.g., H$\beta$) to constrain the intrinsic Lyman-$\alpha$ emission. A simple prior on the emergent Lyman-$\alpha$ emission reduces the number of galaxies required to obtain the same constraints by a factor of about two. These number densities can be achieved with a targeted survey that is complete down to $M_{\rm UV}=-17$ -- -19, depending on the overdensity of the field.

We demonstrated that our framework is not sensitive to the assumed distribution for the emergent Lyman-$\alpha$ emission. 
 We also accurately recovered ionized bubbles when we  applied our framework to 3D EoR simulations.

Our pipeline can be applied to existing observations of Lyman-$\alpha$ spectra from galaxy groups. Additionally, the observational requirements for a statistical detection of local HII regions presented here can be used to design complimentary new {\it JWST} surveys. The application of our framework to multiple independent fields would allow us to constrain the {\it  distribution} of the bubble sizes, which helps us to understand which galaxies reionized the Universe.

\section{Data availability}
The code related to the work is publicly available at \faGithub \href{https://github.com/IvanNikolic21/Lyman-alpha-bubbles}{IvanNikolic21/Lyman-$\alpha$ bubbles}.

\begin{acknowledgements}
      We gratefully acknowledge computational resources of the Center for High Performance Computing (CHPC) at SNS. AM acknowledges support from the Italian Ministry of Universities and Research (MUR) through the PRIN project "Optimal inference from radio images of the epoch of reionization", the PNRR project "Centro Nazionale di Ricerca in High Performance Computing, Big Data e Quantum Computing". CAM and TYL acknowledge support by the VILLUM FONDEN under grant 37459. CAM acknowledges support from the Carlsberg Foundation under grant CF22-1322.
\end{acknowledgements}

\bibliographystyle{aa}
\bibliography{Ly_bub}

\appendix

\section{Mapping the joint distribution over all flux bins with kernel density estimation}
\label{sec:appendixA}
\begin{figure}
    \centering
    \includegraphics[width=\linewidth]{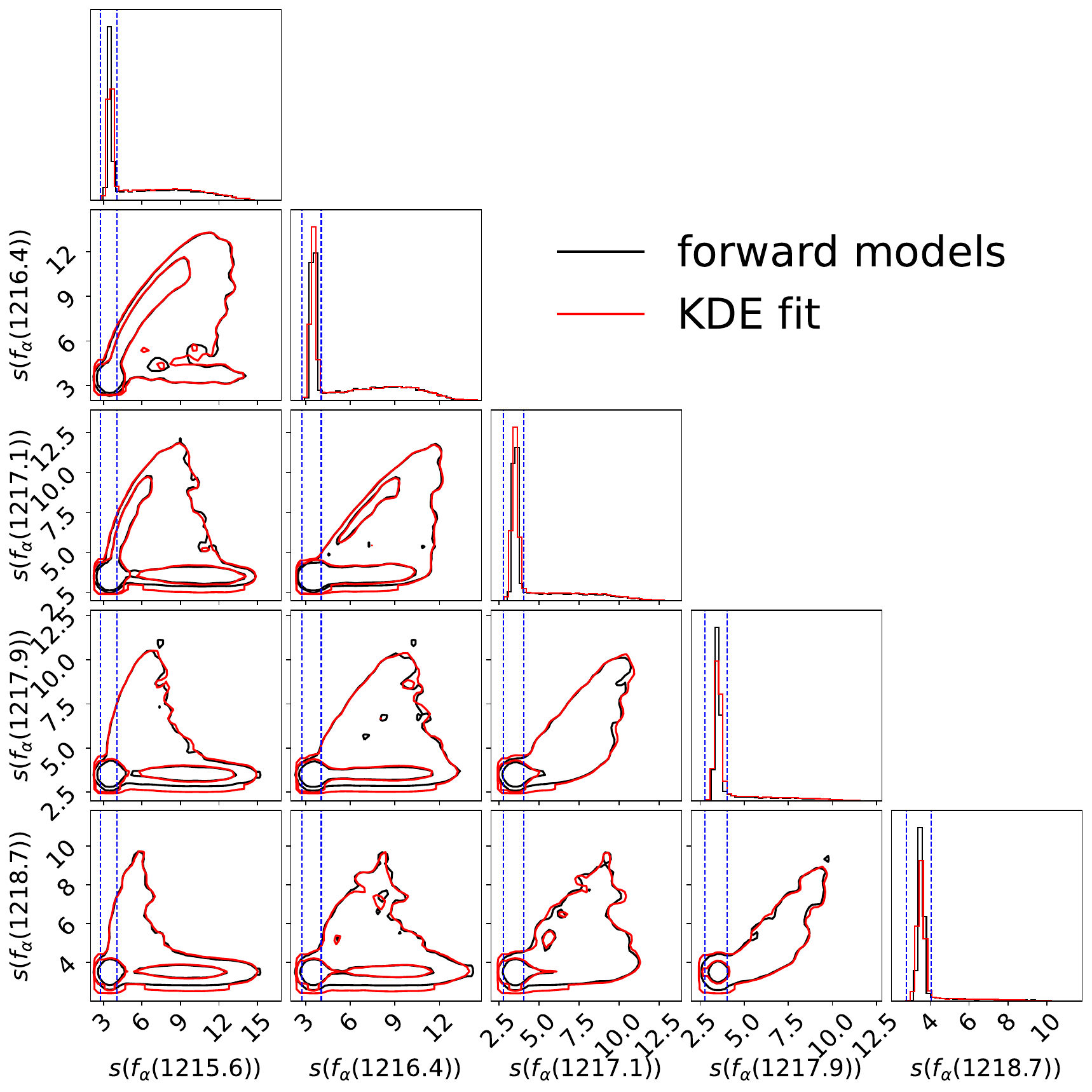}
    \caption{Corner plot of scaled flux spectra and the KDE fit to the forward models. Black contours represent 68\% and 95\% C.L. of the distribution of forward models for a galaxy located in the center of a $R_{\rm b}=10$cMpc ionized bubble. Red contours represent same distributions for the KDE fit to the forward models. Blue dashed lines mark regions with $\pm 2\sigma$ of the noise of the spectrum. Contours outside of these dashed lines represent Lyman-$\alpha$ emission, with non-negligible  correlation between bins.}
    \label{fig:corner flux}
\end{figure}
In Section~\ref{sec:transmissions_more} we showed how we generate forward models of JWST observations. Here we detail how we compute flux PDFs, i.e. likelihoods, in the space of local HII region parameters. In Fig.~\ref{fig:corner flux} in black we show a corner plot of forward models of a galaxy located in the center of its local HII region with size $R_{\rm b}=10$cMpc. Black contours display $68\%$ and $95\%$ C.L. of the distribution respectively. We only show bins in our fiducial resolution ($R~1000$) where the Lyman-$\alpha$ emission is located. In order to fit a density estimator to obtain a smooth PDF, we scale fluxes by a common function:

\begin{equation}
    s(f_{\alpha}(\lambda)) = 93.5 + 5 \log{\left(
    10^{-18} + 2 f_{\alpha}(\lambda)\right)}.
    \label{eq:app1}
\end{equation}

Each column and row represents one wavelength bin of the flux scaled by the function in Eq.~\ref{eq:app1}. We fit a Kernel Density Estimator to the forward models displayed in the corner plot with an exponential kernel of bandwidth $h=0.12$. The scaling function and bandwidth were selected by optimizing the results of Section~\ref{sec:results} (i.e., kernel density bandwidth and normalizing function that allows the inference of bubble properties for the least number of galaxies). We show contours of the fitted KDE in red in Fig.~\ref{fig:corner flux}. 

Several features can be noted in Fig.\ref{fig:corner flux}. Firstly, there is a strong peak for lower values of $s(f_{\alpha}(\lambda))$. This peak corresponds to the noise which is Gaussian by construction. This is further demonstrated with the blue dashed lines that show $5\sigma$ noise levels from Section.~\ref{sec:noise}. On the other hand, there are additional features for larger values of $s(f_{\alpha}(\lambda))$. These features represent the Lyman-$\alpha$ signal that is higher than the noise. It is important to note that there is non-negligible correlation between different bins and the distribution is highly non-trivial. This shows that a simple likelihood that uses independent Gaussians for each bin would potentially bias the inference and an approach that fits the whole distribution is necessary. 

KDE fits the data for all of the bins, in the noise and signal regimes. There is a small overestimation of the width of the distribution for bins that are noise-dominated. Since these bins do not give a lot of information for Lyman-$\alpha$ inference, our KDE is not biasing results.

\end{document}